\documentclass[fleqn,usenatbib]{mnras}

\usepackage[T1]{fontenc}

\DeclareRobustCommand{\VAN}[3]{#2}
\let\VANthebibliography\thebibliography
\def\thebibliography{\DeclareRobustCommand{\VAN}[3]{##3}\VANthebibliography}

\usepackage{graphicx}	
\usepackage{tabularx}
\usepackage{amsmath}	
\usepackage{amssymb}	
\usepackage{orcidlink}

\usepackage{newtxtext,newtxmath}

\newcommand{\bs}{\boldsymbol}

\usepackage{xcolor}

\title[A View of the Galactic Bar with {\it Gaia} LPV]{Kinematics and dynamics of the Galactic bar revealed by {\it Gaia} long-period variables}

\author[Zhang et al.]{Hanyuan Zhang$^{1}$\thanks{hz420@cam.ac.uk (HZ)}\orcidlink{0009-0005-6898-0927},
Vasily Belokurov$^{1}$\orcidlink{0000-0002-0038-9584}, N. Wyn Evans$^{1}$, Sarah G. Kane$^{1}$\orcidlink{0000-0001-8411-1012} and\newauthor Jason L. Sanders$^{2}$
\orcidlink{0000-0003-4593-6788}\\
\\
$^{1}$ Institute of Astronomy, University of Cambridge, Madingley Road, Cambridge CB3 0HA, UK \\
$^{2}$ Department of Physics and Astronomy, University College London, London WC1E 6BT, UK\\
}

\date{Accepted XXX. Received YYY; in original form ZZZ}

\pubyear{2024}

\begin{document}
\label{firstpage}
\pagerange{\pageref{firstpage}--\pageref{lastpage}}
\maketitle

\begin{abstract}
We use low-amplitude, long period variable (LA-LPV) candidates in \textit{Gaia} DR3 to trace the kinematics and dynamics of the Milky Way bar. LA-LPVs, like other LPVs, are intrinsically bright and follow a tight period-luminosity relation, but unlike e.g. Mira variables, their radial velocity measurements are reliable due to their smaller pulsation amplitudes. We supplement the \textit{Gaia} astrometric and radial velocity measurements with distance moduli assigned using a period-luminosity relation to acquire full 6D phase space information. The assigned distances are validated by comparing to geometric distances and StarHorse distances, which shows biases less than $\sim5\%$. Our sample provides an unprecedented panoramic picture of the inner Galaxy with minimal selection effects. We map the kinematics of the inner Milky Way and find a significant kinematic signature corresponding to the Galactic bar. We measure the pattern speed of the Galactic bar using the continuity equation and find $\Omega_{\rm b}=34.1\pm2.4$~km s$^{-1}$ kpc$^{-1}$. We develop a simple, robust and potential-independent method to measure the dynamical length of the bar using only kinematics and find $R_{\rm b}\sim4.0$~kpc. We validate both measurements using N-body simulations. Assuming knowledge of the gravitational potential of the inner Milky Way, we analyse the orbital structure of the Galactic bar using orbital frequency ratios. The $x_1$ orbits are the dominant bar-supporting orbital family in our sample. Amongst the selected bar stars, the $x_1 v_1$ or "banana" orbits constitute a larger fraction ($\sim 15\%$) than other orbital families in the bar, implying that they are the dominant family contributing to the Galactic X-shape, although contributions from other orbital families are also present. 
\end{abstract}




\section{Introduction}

The Milky Way hosts an elongated, non-axisymmetric bar at the centre, a common feature for disc galaxies. The Milky Way bar was first observed in non-axisymmetric gas flow \citep{Peters_1975, Binney_1991}, near-infrared emission \citep{Blitz_1991} and star counts \citep{Nakada_1991, Stanek_1997}. A revolution in the understanding of the Galactic bar has occurred in the past two decades using red clump stars as standard candles due to their narrow luminosity function \citep{McWilliam_Zoccali_2010, Nataf_2010}. Without resolving the locations of individual stars, the distribution of red clump stars can be inferred statistically by inverting the observed luminosity function along each line-of-sight \citep[see][]{Wegg_Gerhard_2013}. 3D density maps of the inner Galaxy are constructed using this technique \citep{Wegg_Gerhard_2013, Wegg_2015, Simion_2017, Sanders_2019a, Paterson_2020}.
Augmenting the 3D density distribution of the inner Galaxy with kinematics from the BRAVA and ARGOS surveys, \citet{Portail_2015b, Portail_2017} built dynamical models and found the effective potential of the Galactic bar using the made-to-measure (M2M) method. Orbital structures of the Galactic bar in the M2M models were analysed in \citet{Portail_2015}.

The pattern speed of the Galactic bar is a crucial parameter because it sets the resonance radii for stars and affects the kinematics in the Solar neighbourhood \citep{Binney_2020, Kawata_2021}, in the disc \citep{Chiba_2021, Gaia_Collaboration_dr3disc} and in the halo \citep{Davies_2023, Dillamore_2023, Dillamore_2024}. Older measurements of gas kinematics reported a fast-short bar with a pattern speed of $50-60$~km s$^{-1}$ kpc$^{-1}$ \citep{Fux_1999, Englmaier_1999}, while more recent measurements of gas give a smaller value of $40$~km s$^{-1}$ kpc$^{-1}$ \citep[e.g][]{Sormani_2015, Li_2022}. Compared to the relatively fast pattern speed measured from gas kinematics, observations of the resonance features in the Solar neighbourhood give a slower pattern speed in the range of $34-40$~km s$^{-1}$ kpc$^{-1}$ \citep{Binney_2020, Chiba_2021, Kawata_2021, Gaia_Collaboration_dr3disc, Dillamore_2023, Dillamore_2024}. Directly observing the central region of the Galaxy, \citet{Clarke_2022} measured a pattern speed of $\sim33$~km s$^{-1}$ kpc$^{-1}$ using M2M method with \textit{Gaia} and VIRAC proper motion measurements. Applying the continuity equations to the stars at the inner Milky Way yielded a pattern speed of $\sim41$~km s$^{-1}$ kpc$^{-1}$ \citep{Sanders_2019, Bovy_2019, Leung_2023}. 

Parallax measurements from {\it Gaia} are usually not sufficiently precise in the region of the Galactic bar \citep{BJ_2021}. Previous attempts to acquire full 6D phase space measurements have relied on the spectroscopy-based distances \citep[\texttt{StarHorse} and \texttt{AstroNN}, ][]{Bovy_2019, Queiroz_2020, Queiroz_2021, Arentsen_2024}, which therefore have strong selection functions. The heavy dust extinction also induces an additional selection bias in the inner Galaxy, further complicating studies of the Galactic bar. In this work, we instead use the period-luminosity relations (PLR) of long period variables (LPV) to assign distances. Structures in the Milky Way have been intensively studied using variables as tracers because of their tight PLR, most famously RR Lyrae \citep{Iorio2018,Prudil_2022, Semczuk_2022} and Cepheid variables \citep{Lemasle_2022, Matsunaga_2023}. However, both are biased tracers -- RR Lyrae are typically old \citep[but see][]{Bobryck2024}, while (classical) Cepheids are limited to young ages of $<1$ Gyr. LPVs, as another type of pulsating stars, exhibit pulsations with longer periods, ranging from $10$~days to above $1000$. LPVs follow tight PLRs and can be used as distance indicators \citep[see e.g.][]{Rau_2019, Trabucchi_2021, Sanders_2023}. Also, because LPVs are redder than RR Lyrae and Cepheids, they are less affected by dust extinction. Hence, LPVs are popularly used to trace the structure of the Milky Way \citep{Catchpole_2016, Grady_2020, Sanders_2022, Iwanek_2023, Hey_2023, Sanders_2024}. Among several types of LPVs, Mira variables have a unique property wherein their period and age are correlated \citep{FeastWhitelock2000, Grady_2019, Trabucchi_2022, Zhang_2023}. The ages of Mira variables range from $\sim2$ to $12$~Gyr, implying that these LPVs can be used as an unbiased tracer of the entire wide mix of stellar populations of the Galaxy. Furethermore, the period-age relation in \citet{Zhang_2023} can be used to date the formation epoch of the Galactic bar \citep{Sanders_2024}. 

In this paper, we use OSARGs \citep[OGLE Small Amplitude Red Giants,][]{Wray_2004_OSARG} instead of Mira variables as tracers. In this paper, we refer to these as low-amplitude, long-period variables (LA-LPV). LA-LPVs have characteristic periods between 0 to over 100 days and are more abundant than Miras. They serve as distance indicators with uncertainties smaller than $\sim 15\%$ \citep{Rau_2019,Hey_2023}. Light curves of LA-LPVs show multi-periodic behaviour and small amplitude fluctuations with $G$-band amplitude$\lesssim0.2$ \citep{Soszynski_2004}. They are distinguished from other LPVs by their distinct PLRs \citep{Wood_2000, OGLELMC_2007}: they have smaller periods and fainter brightness compared to Miras and semi-regular variables (SRV). Pragmatically, LA-LPVs are better kinematic tracers than Mira variables because their pulsation amplitudes are smaller, so radial velocity measurements from \textit{Gaia} remain useful.
The OGLE survey \citep{OGLE} discovered and investigated OSARGs in detail with selected candidates in the LMC, SMC \citep{OGLELMC_2009} and the Galactic bulge \citep{OGLEBLG}. \citet{Hey_2023} used LA-LPV and SRV candidates in the OGLE-BLG survey \citep{OGLEBLG} to investigate the kinematics of the Galactic bar. They found a clear quadrupole pattern in the velocity field corresponding to the bar, but the presence of a selection function caused by the OGLE footprint made further quantitative analysis difficult. In this work, we explore the \textit{Gaia} LPV catalogue \citep{GaiaDR3_variable, GaiaDR3_LPV}. Thanks to the full-sky coverage of \textit{Gaia}, there is no sharp truncation in our LA-LPV sample compared to the sample in OGLE-BLG \citep{Hey_2023}. Therefore, we can build a panoramic view of the inner Galaxy with resolved stars using the full 6D phase space, which allows us to study the kinematics and dynamics of the Galactic bar in detail.

This paper is structured as follows: we present the catalogue construction and distance assignment in section~\ref{sec::data}. In section~\ref{sec::kinematics}, we map the kinematics of the Galactic centre and compare our observational results with simulations. We also quantify the pattern speed and bar length solely using stellar kinematics. We investigate the dynamical properties and orbital structures of the LPV tracers in section~\ref{sec::dynamics}. In section~\ref{sec::discussion}, we discuss and compare our pattern speed and bar length measurements with previous studies. We conclude our analysis in section~\ref{sec::conclusion}.

\section{Data}
\label{sec::data}
\subsection{Gaia Long Period Variables}

\begin{figure*}
    \centering
    \includegraphics[width = 1.7\columnwidth]{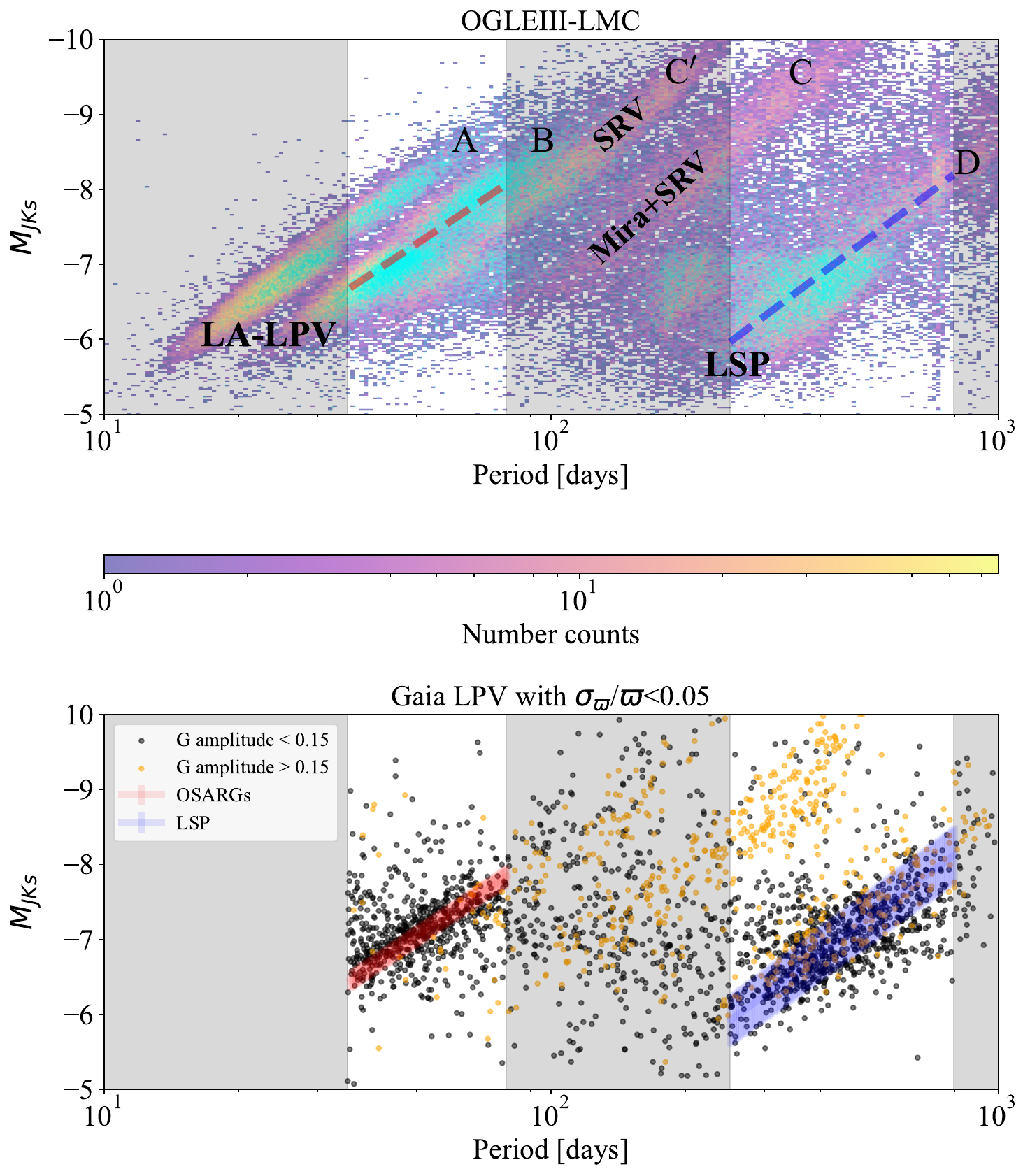}
    \caption{Period-luminosity plane for LPVs. {\it Top:} Background density in the $P-M_{\mathrm{JKs}}$ plane for all LPV candidates in OGLEIII-LMC. Here each star has up to three reported periods together with the associated amplitudes, so each star can contribute three times to the 2D histogram. The foreground aqua dots are stars with pulsation modes with $I$ amplitude between $0.02$ and $0.06$. These illustrate the power of amplitude cuts on cleaning the $P-M_{\mathrm{JKs}}$ sequences. {\it Bottom:} As above but for the {\it Gaia} LPV catalogue with $\sigma_\varpi/\varpi<0.05$. For these stars the distance modulus is computed using the geometric distance from \citet{BJ_2021}. {\it Gaia} LPV candidates are coloured according to their reported amplitude, where stars with $G$ amplitude smaller than $0.15$ are black, and all other stars are orange. $G$ amplitude is the amplitude reported in \textit{Gaia} LPV catalogue which corresponds to the half peak-to-peak Fourier amplitude in the $G$-band. The red and blue lines are the fitted OSARG and LSP period-luminosity sequences, where the width of the line represents the $1\sigma$ intrinsic scatter in the sequence. In both panels, we shade in grey the period ranges that we exclude when constructing the final sample.}
    \label{fig::OGLE_Gaia_PL_plane}
\end{figure*}

We use the specific object study (SOS) table from the LPV catalogue of \textit{Gaia} DR3 \citep{GaiaDR3_LPV, Gaia_DR3} to select LA-LPV candidates. This catalogue is constructed using supervised classification of identified variable stars in \textit{Gaia} data with features including parallax, colours, light curve statistics and Lomb-Scargle period \citep{GaiaDR3_variable}. Stars classified as LPVs that have the 5th-95th percentile of their G-band light curves greater than 0.1 mag and $G_{BP}-G_{RP} > 0.5$ (together with some other minor cleaning cuts) enter the final SOS table \citep[see details in][]{GaiaDR3_LPV}. \cite{GaiaDR3_variable} compared the SOS LPV catalogue with the ASAS-SN \citep{ASAS_SN} survey and the OGLE \citep{OGLE} survey and found a contamination of $\sim 1 \%$. 

The periods of the identified LPVs in the SOS table are determined from the $G$ light curves using the generalised Lomb-Scargle periodogram \citep{VanderPlas_2018}. Periods are only published if the primary period is greater than 35 days and smaller than 34 months (the duration of the time series). The amplitude corresponds to the half peak-to-peak amplitude of the fundamental Fourier mode, which we refer to as the $G$ amplitude in the later discussion. Due to the amplitude and period cut in the \textit{Gaia} LPV catalogue, small amplitude LA-LPVs on the fainter magnitude end are excluded.

We acquire infrared photometry of the {\it Gaia} LPV candidates from the 2MASS catalogue \citep{2MASS} using a cross-match with $1''$ radius. Removing stars with low-quality $JHK_s$ measurements by requiring $\texttt{ph\_qual=AAA}$ and $\texttt{gal\_contam}=0$, we retain $131,561$ LPV candidates with valid $J$ and $K_s$ band photometry. We compute the colour-corrected Wesenheit index $W_{\mathrm{JKs}}$ using
\begin{equation}
    W_{\mathrm{JKs}} = K_\mathrm{s} - 0.686\times(J-K_\mathrm{s}),
\end{equation}
and we define the absolute magnitude corresponding to $W_{\mathrm{JKs}}$ as $M_{\mathrm{JKs}}$ \citep{Cardelli_1989, OGLELMC_2007}. Although the extinction law at the inner Galaxy is flatter than the one used in \citet{Cardelli_1989}, we treat this as a second-order effect \citep{Nishiyama_2009}.

\subsection{Selection of LA-LPV candidates in the {\it Gaia} LPV catalogue}

We start by selecting stars sitting on the period-luminosity ridges of OSARGs \citep[sequence A and B in ][]{Wood_2000} and Long Secondary Period (LSP, sequence D). The physical origin of these pulsation sequences is still under debate, but for our purposes all that matters is that they are empirically secure. 
OSARGs are characterised by their small amplitudes (typically $G$ amplitude $\lesssim0.2$~mag) and multi-periodic behaviour. The periods of OSARGs are mainly in the range of $10$ to $100$~days; however, many OSARGs also have characteristic periods longer than 200 days associated with the LSP. \cite{Rau_2019} demonstrated the correlation between the period, luminosity and amplitude of OSARGs. Because the \textit{Gaia} LPV catalogue only provides the primary period and corresponding amplitude, selecting OSARGs using their multi-periodic signatures from this catalogue is challenging. Hence, we turn to the period and amplitude features. In the top panel of Fig.~\ref{fig::OGLE_Gaia_PL_plane}, we plot LPV candidates in the OGLEIII-LMC survey \citep{OGLELMC_2009} in the period-luminosity ($\mathrm{M_{\mathrm{JKs}}}$) plane. LPV candidates with $I$ amplitudes of the pulsation mode between $0.02$ and $0.06$ are overlaid as aqua dots, and the 2D histogram of the number counts of all LPV candidates (amplitude range spans from $0.01$ to $\sim0.2$) is plotted in the background. The period-$M_{\mathrm{JKs}}$ relation becomes less complicated after cutting $0.02<I\,\mathrm{amplitude}<0.06$, as a result of which only three prominent sequences are left. The two main ridges corresponding to the main OSARG and LSP period-$M_{\mathrm{JKs}}$ relations are denoted by the red and blue dashed lines, which are the PLR calibrated in \citet{OGLELMC_2007}. For now, we neglect the secondary OSARG sequence on the leftmost side of the plane (sequence A) as it is a subdominant population. There is a noticeable horizontal gap in the OGLEIII-LMC OSARG candidates around $M_{\mathrm{JKs}}\sim-7.2$ likely due to the saturation of the bright stars.

Inspecting the period-$M_{\mathrm{JKs}}$ sequences of LPVs in the Milky Way, we select stars with high-quality parallax using fractional parallax uncertainty, $\sigma_\varpi/\varpi$, $<0.05$, and we plot them on the period-$M_{\mathrm{JKs}}$ plane as shown in the lower panel of Fig.~\ref{fig::OGLE_Gaia_PL_plane}, where the distance moduli are computed using geometric distances from \cite{BJ_2021}. We find multiple sequences in the window for the LSP period-luminosity sequence due to the presence of SRVs and Mira variables. However, different sequences can be well distinguished using the amplitude of the pulsation mode, as SRV and Mira variables have larger $G$ amplitudes than OSARGs \citep{OGLELMC_2007, Matsunaga_2009, Grady_2019}. LPVs with $G$ amplitude $<0.15$ are shown by the black dots, and on those two unshaded regions, the period-luminosity sequences of OSARGs and LSPs are conspicuous and denoted by the red and blue solid lines. Therefore, we remove stars with $G\,\mathrm{amplitude}>0.15$ to isolate LA-LPV candidates from the \textit{Gaia} LPV catalogue.

We further clean the LA-LPV candidates from young stellar object (YSO) contamination, as YSOs share similar colours, amplitudes and periods with LPVs \citep{Mowlavi_2018}. \cite{Zhang_2023} showed that the $\texttt{best\_class\_score}$, the probability of the object being the reported class from the classification pipeline, is useful to remove potential YSO contamination. Hence, we cut using $\texttt{best\_class\_score}>0.8$. We also remove stars with primary periods between $1.9<\log_{10}(\mathrm{Period/days})<2.4$ and $\log_{10}(\mathrm{Period/days})>2.9$ to clean LA-LPV stars that are not sitting on OSARG or LSP period-luminosity tracks as shown by the grey shaded region in Fig.~\ref{fig::OGLE_Gaia_PL_plane}.

\subsection{Distance Assignment using the Period-luminosity Relation}

\begin{figure}
    \centering
    \includegraphics[width = \columnwidth]{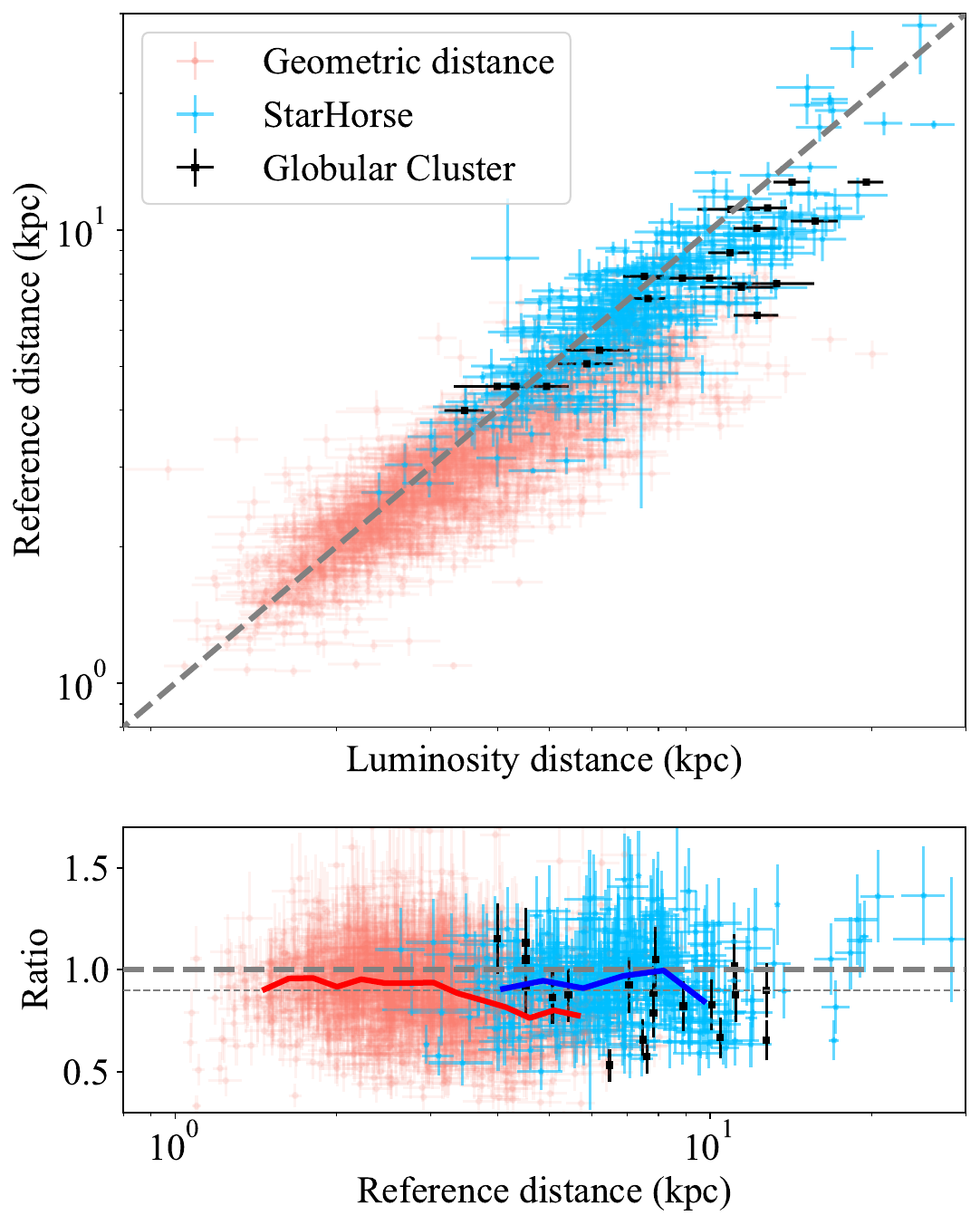}
    \caption{Distance validation. Comparison between the luminosity distances computed in this work and the literature distances for stars in common. Parallax-based geometric distances from \citet{BJ_2021} are shown in red, spectroscopic \texttt{StarHorse} distances from \citet{Queiroz_2020} in blue, and ditances to globular cluster members from \citet{Vasiliev_2021b} are shown as black dots. {\it Top:}  Literature values from the catalogues above are on the $y$-axis and our luminosity distances on the $x$-axis. The grey dashed line represents the 1:1 relation. {\it Bottom:} Ratio of the reference distance and the luminosity distance as a function of the reference distance. The thick grey horizontal line is ratio of 1. The thinner dotted horizontal line shows the $90\%$ consistency (1:0.9 ratio). The red (blue) line shows the median distance ratio for the geometric (\texttt{StarHorse}) distances as reference. We argue that our luminosity distance estimates are $\sim 95\%$ consistent with the references considered.}
    \label{fig::Distance_crossvalidation}
\end{figure}

\begin{figure*}
    \includegraphics[width = 0.96\textwidth]{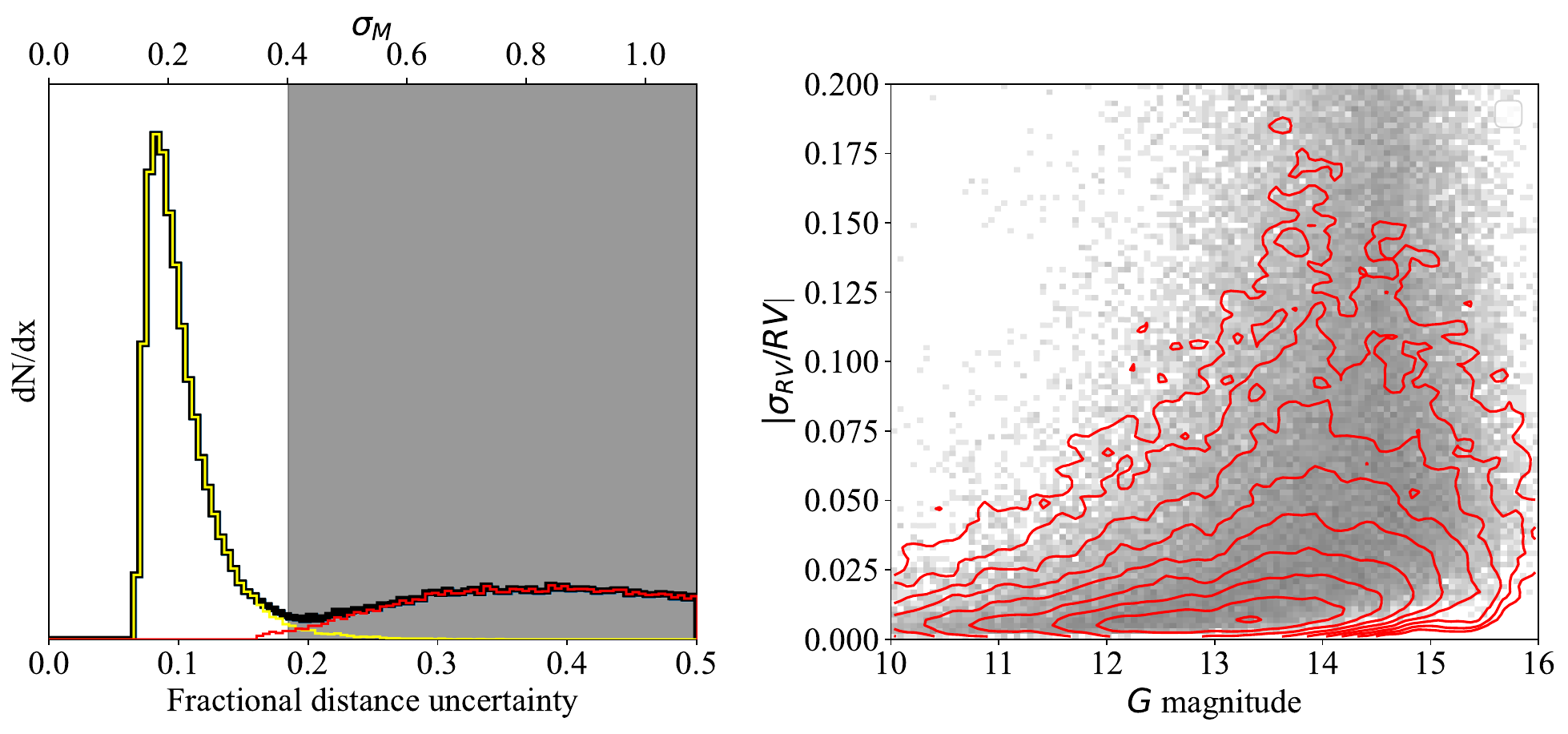}
    \caption{{\it Left:} Fractional distance uncertainty distribution of our sample. The corresponding uncertainty in the absolute magnitude is shown in the top x-axis. The yellow and red lines are the distance uncertainty for OSARG and LSP variables, and the black line is the distance uncertainty distribution for the sample overall. We cut the final sample at $\sigma_M<0.4$, corresponding to the $\sim18\%$ distance uncertainty. Grey masked region shows the stars removed from the sample due to their large uncertainties. {\it Right:} Fractional radial velocity uncertainty in our final sample (red contours) compared to a randomly selected {\it Gaia} sample (black 2D histogram). The distribution of the {\it Gaia} radial velocity measurements for our LA-LPVs matches closely the distribution of the randomly selected {\it Gaia} sample. Thus, we argue, the radial velocities of LA-LPVs show no signs of being affected by variability.}
    \label{fig::sample_info}
\end{figure*}

\begin{figure}
    \centering
    \includegraphics[width = \columnwidth]{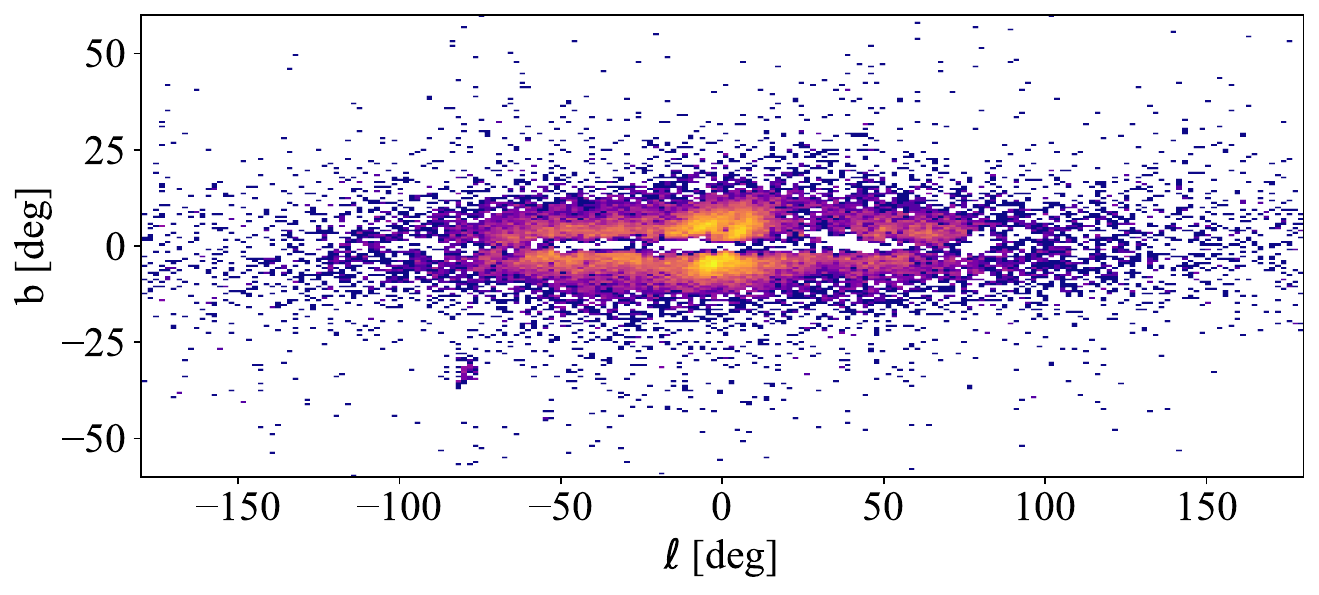}
    \includegraphics[width = \columnwidth]{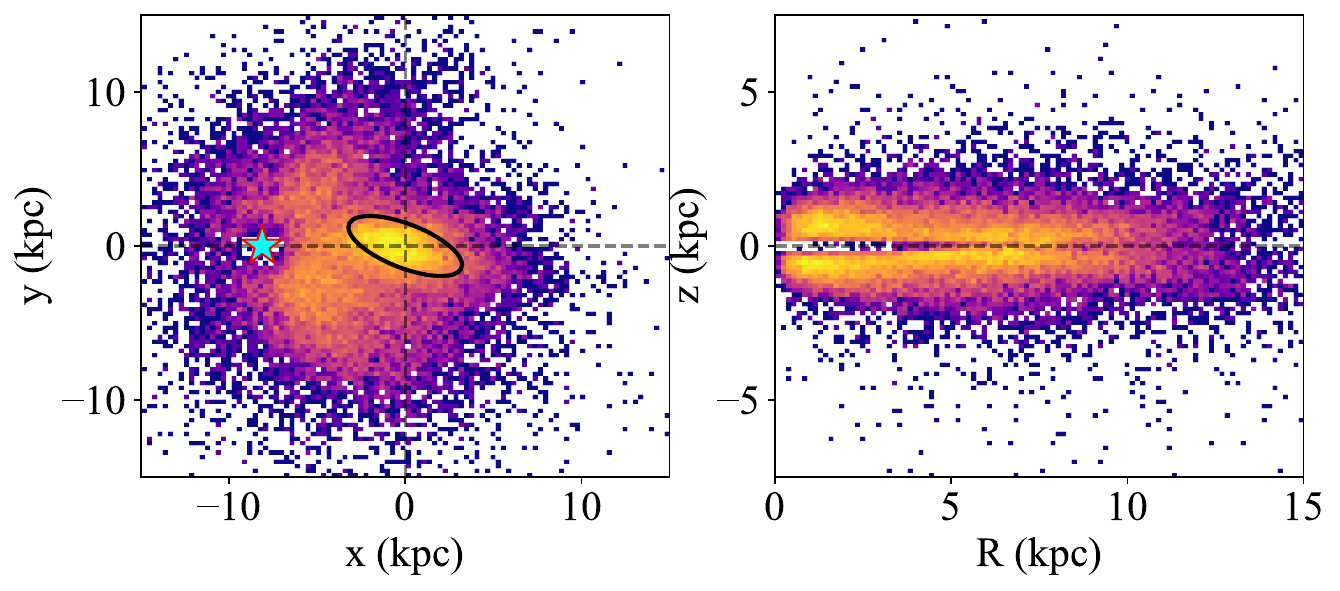}
    \caption{Spatial properties of the final sample. {\it Top:} Distribution of stars in the final sample in Galactic longitude and latitude, $\ell$ and $b$. {\it Bottom left:} Top view of the sample in Galactocentric Cartesian coordinates. The Sun is located at $x = -8.122$~kpc and is labelled by the aqua star. The Galactic bar is marked with the black ellipse and has a viewing angle (between the ellipse's major axis and the line connecting the Sun and the Galactic Centre) of  $25^\circ$. {\it Bottom right:} The Galactocentric cylindrical distribution of the sample.}
    \label{fig::LB&XY&RZ}
\end{figure}

Due to the sparse sampling of the \textit{Gaia} light curves, the period uncertainties are significantly larger than those obtained with OGLE data. The intrinsic scatter in the PLR, the period uncertainty, and the magnitude scatter from single-epoch 2MASS photometry all broaden the width of the period-$M_{\mathrm{JKs}}$ sequence in {\it Gaia} data. Hence, instead of calibrating the period-$M_{\mathrm{JKs}}$ relation of OSARGs and LSPs from the {\it Gaia} LPV catalogue, we use the parametrised period-$M_{\mathrm{JKs}}$ relation calibrated in \citet{OGLELMC_2007} and treat the uncertainty on the fitted parameters therein as the intrinsic scatter of the period-luminosity relation. \cite{OGLELMC_2007} fitted $W_{\mathrm{JKs}}$ as a linear function of $\log(P)$ in the form
\begin{equation}
    W_{\mathrm{JKs}} = \alpha \log(P) + \beta
    \label{eqn::PL_relation}
\end{equation}
where $P$ is the period of the pulsation mode. We convert $W_{\mathrm{JKs}}$ to $M_{\mathrm{JKs}}$ using the LMC distance modulus of $18.48$ \citep{Pietrzynski_2019}. As shown in \citet{Tabur_2010}, the period-luminosity sequence of OSARGs shifts slightly when surveys use different amplitude cuts due to the correlation in the period-luminosity and amplitude. Therefore, to adapt the amplitude cut in the \textit{Gaia} LPV catalogue (and also the Milky Way environment), we further introduce an offset and fit the offset using stars in \textit{Gaia} LPV with $\varpi/\sigma_\varpi>20$, i.e. re-calibrate the P-$M_{\mathrm{JKs}}$ relation using black dots in the lower panel of Fig.~\ref{fig::OGLE_Gaia_PL_plane}. The adapted parameters in Eq.~\ref{eqn::PL_relation} are $\alpha_{\mathrm{OSARG}} = -3.96\pm0.04$ and $\beta_{\mathrm{OSARG}} = -17.95\pm0.13$; $\alpha_{\mathrm{LSP}} = -4.40\pm0.11$ and $\beta_{\mathrm{LSP}} = -23.41\pm0.23$. Compared to the PLR in \citet{OGLELMC_2007}, both OSARGs and LSP sequences have offsets on the order of $\sim 0.2$~mag. The discussion of the exact causes of this offset is beyond the scope of this work. These two sequences are shown as the red and blue thick lines in the lower panel of Fig.~\ref{fig::OGLE_Gaia_PL_plane}, where the width of the lines shows $1\sigma$ scatter of the P-$M_{\mathrm{JKs}}$ relation when the period uncertainty is $0$. We assign $M_{\mathrm{JKs}}$ to stars with the \textit{Gaia} period between $1.6<\log(P/\mathrm{day})<1.9$ using $\alpha_{\mathrm{OSARG}}$ and $\beta_{\mathrm{OSARG}}$ and stars with $2.4<\log(P/\mathrm{day})<2.9$ using $\alpha_{\mathrm{LSP}}$ and $\beta_{\mathrm{LSP}}$. We subsequently calculate the distance moduli and the luminosity distances. The uncertainty of the luminosity distances is caused by the propagated \textit{Gaia} period uncertainty and the intrinsic scatter of the $P-M_{\mathrm{JKs}}$ relation using the uncertainty in the fitted PLR. We neglect the difference between O-rich and C-rich OSARGs, as there is no obvious distinction in their $P-M_{\mathrm{JKs}}$ sequences, and the majority ($\sim97\%$) of stars in our sample are O-rich LPVs according to the $\mathrm{\texttt{is\_cstar}}$ label in the \textit{Gaia} LPV catalogue \citep{GaiaDR3_LPV}. \citet{Sanders_Matsunaga_2023} show that the $\mathrm{\texttt{is\_cstar}}$ label classifies highly extincted stars as C-rich, which means the true fraction of O-rich stars is even higher than $97\%$. We have also inspected the metallicity dependence of the period-$M_{\mathrm{JKs}}$ relation and treat this as a second-order effect (Zhang et al. in prep.). 

We cross-validate the luminosity distances with various catalogues that have distances, including parallax-based geometric distances \citep{BJ_2021}, spectroscopic \texttt{StarHorse} distances \citep{Queiroz_2020}, and distances to globular cluster members \citep{Vasiliev_2021b}. For the geometric distances, we cut the comparison sample to select stars with small fractional parallax uncertainties, i.e. $\sigma_\varpi/\varpi<0.1$. We focus on the comparison with the geometric distances and \texttt{StarHorse} distances as they have a sufficient number of cross-matches with our data. Stars with distance modulus uncertainties greater than $0.4$ (roughly corresponding to $\sim20\%$ distance uncertainties) are removed from our sample, which removes stars with large period uncertainties. In Fig.~\ref{fig::Distance_crossvalidation}, we plot the luminosity distances versus the reference distances from the other catalogues in the upper panel and the ratio between the two distances versus the reference distances in the lower panel. The grey dashed lines represent the 1:1 line. Overall, we demonstrate a good 1:1 correlation between the assigned luminosity distances and the distances calibrated using various other methods. For the geometric and \texttt{StarHorse} distances, we take bins of the reference distances and compute the median of the distance ratio in each bin. We plot the median of the distance ratios as the red and blue lines, respectively, for the geometric and \texttt{StarHorse} distances. For the comparison with the geometric distances, the median ratio is consistent with the 1:1 lines at small distances and deviates at large distances, as expected for the parallax-based distances. The lower panel of Fig.~\ref{fig::Distance_crossvalidation} reveals that the luminosity distances agree with the other distance calibrations with only $\sim5\%$ deviation.

\subsection{Final sample}

We summarise the quality cuts applied to the \textit{Gaia} LPV catalogue below. We cross-match \textit{Gaia} LPV SOS table to 2MASS for infrared photometry, and we remove stars without good $JHK_{\rm s}$ photometry  by requiring $\texttt{ph\_qual}=\mathrm{AAA}$ and remove galaxy contamination by requiring $\texttt{gal\_contam}=0$. We remove potential YSO contamination using $\texttt{best\_class\_score}>0.8$. We retain stars only if the \textit{Gaia} published period is between $1.6<\log(P\mathrm{[days]})<1.9$ or $2.4<\log(P\mathrm{[days]})<2.9$ and $G\,\mathrm{amplitude}<0.15$ to select LA-LPV candidates. After assigning the luminosity distances, we remove stars with magnitude uncertainties greater than $0.4$, i.e. $\sigma_{M}<0.4$, corresponding to cutting at $\sim20\%$ fractional distance uncertainty. This effectively cuts on the period uncertainty as the intrinsic scatter of the $P-M_{\mathrm{JKs}}$ relation is fixed. The published \textit{Gaia} period uncertainties associated with the LSP sequence are large, and thus, only a few stars on that sequence survive the magnitude uncertainty cut. Over $98\%$ of stars in the final sample have their luminosity distances assigned using OSARG's period-luminosity sequence because the period uncertainties associated with LSP candidates are large. There are 45,819 stars left. In the left panel of Fig.~\ref{fig::sample_info}, we show the distribution of the fractional distance uncertainties of this sample. The proper motion uncertainty is $\sim2-5\%$, and the mean and median distance uncertainty of this sample is $\sim10\%$. Hence, the distance uncertainty is the dominate source of uncertainty. To perform kinematic and dynamical analysis, we further remove stars without radial velocity measurements from {\it Gaia} DR3. This leaves us with a final sample size of 33\,704 stars with full 6-D phase space measurements. 

To validate the radial velocity measurements for this LA-LPV sample, we compare the fractional radial velocity uncertainty, $|\mathrm{\sigma_{RV}/RV}|$, to a randomly selected \textit{Gaia} sample with radial velocity measurements. We randomly select 100\,000 stars from \textit{Gaia} with radial velocity measurements, and we cross-match it to the sample constructed above and remove common stars. In the right panel of Fig.~\ref{fig::sample_info}, we plot the fractional radial velocity uncertainty against $G$ magnitude. The background shows the distribution of the randomly selected \textit{Gaia} sample, and the red contours are the distribution of our sample. The Figure demonstrates that the small variability of stars in this sample does not affect the radial velocity measurement from \textit{Gaia} as both samples have very similar distributions.

Fig.~\ref{fig::LB&XY&RZ} shows the on-sky distribution and the spatial coverage of the final sample in Galactic celestial, Galacto-centric Cartesian and cylindrical coordinates (see Section~\ref{sec::kinematics} for details). Most stars reside below $|b|\sim25^\circ$ and $|z|\sim2.5$ kpc, where $R$ and $z$ are the Galactocentric cylindrical coordinates. The sample traces the Milky Way from the centre out to $R\sim15$~kpc, showing complete coverage of the inner Milky Way. Some of these stars have large luminosity distances extending to the other side of the Galaxy. The face-on view of the Galactic disc shows an overdensity compatible with the shape and orientation of the Galactic bar, giving us a unique opportunity to view the entire bar and study the kinematics of stars on the other side of the bulge. 

\subsection{Selection effects}
Because our selection of LA-LPV candidates is based purely on photometry and given the full-sky coverage of \textit{Gaia}, the LA-LPV sample produces no sharp truncation in the spatial map. This allows us to present a panoramic map of the inner Galaxy and study the kinematics and dynamics homogeneously. The LA-LPV and SRV candidates in OGLE-BLG compiled by \citet{OGLEBLG} and analysed in \citet{Hey_2023} have a sample size larger than our dataset because OGLE reaches fainter magnitudes than Gaia and has a higher recovery rate of LPV candidates due to higher temporal cadence. However, due to the limited footprint of the OGLE-BLG survey \citep{OGLEBLG}, its selection function has a large-amplitude spatial variation. Compared to OGLE, except for the incompleteness in star counts within $|b|<2^\circ$ due to the dust extinction as shown in the top panel of Fig.~\ref{fig::LB&XY&RZ}, in our sample, the observational footprint does not impose extra spatial variation in the selection function.

In addition to the selection function caused by the extinction, there is a selection function in the apparent magnitude caused by the radial velocity cut. To the first order, this can be approximated as an absolute magnitude cut that changes with the distance from the Sun. However, due to the high intrinsic luminosity of LPVs ($M_{\mathrm{JKs}}\lesssim-6$) as well as the period and amplitude cuts applied, we limit the absolute magnitude to $-8\lesssim M_{\mathrm{JKs}}\lesssim -7$ for stars in our sample. This limit has the effect that a similar population of stars is selected at all heliocentric distances, except for stars that are very close (heliocentric distance $\lesssim2$~kpc, where they are saturated) or very far (heliocentric distance $\gtrsim12$~kpc, where stars are too faint to have radial velocity measurements). Therefore, the selection function in magnitude does not cause spatial variation on the number of selected stars as a function of heliocentric distances, to the first order.

\section{Kinematics of the Galactic bar}
\label{sec::kinematics}

To focus on the inner Galaxy, we only keep stars with $|x|<5$~kpc and $|y|<5$~kpc. With the full 6D phase space measurements of $\sim$20\,000 stars, we start by mapping the kinematics in the central Milky Way. We derive the positions and velocities in Galactocentric Cartesian coordinates and right-handed  Galactocentric cylindrical coordinates from astrometric and radial velocity measurements from {\it Gaia} DR3 and use the assigned luminosity distances instead of parallax. The location of the Sun is $x = -8.122$~kpc, and the solar velocities are ($12.9$, $245.6$, $7.78$)~km/s in the radial, azimuthal, and vertical directions, respectively \citep{Schonrich_2010}. The uncertainties are propagated using the Monte Carlo method. In this section, we analyse the kinematics of this sample and use N-body simulation results as references to calibrate the parameters of the Galactic bar.

\subsection{Velocity fields}

\begin{figure*}
    \includegraphics[width = \textwidth]{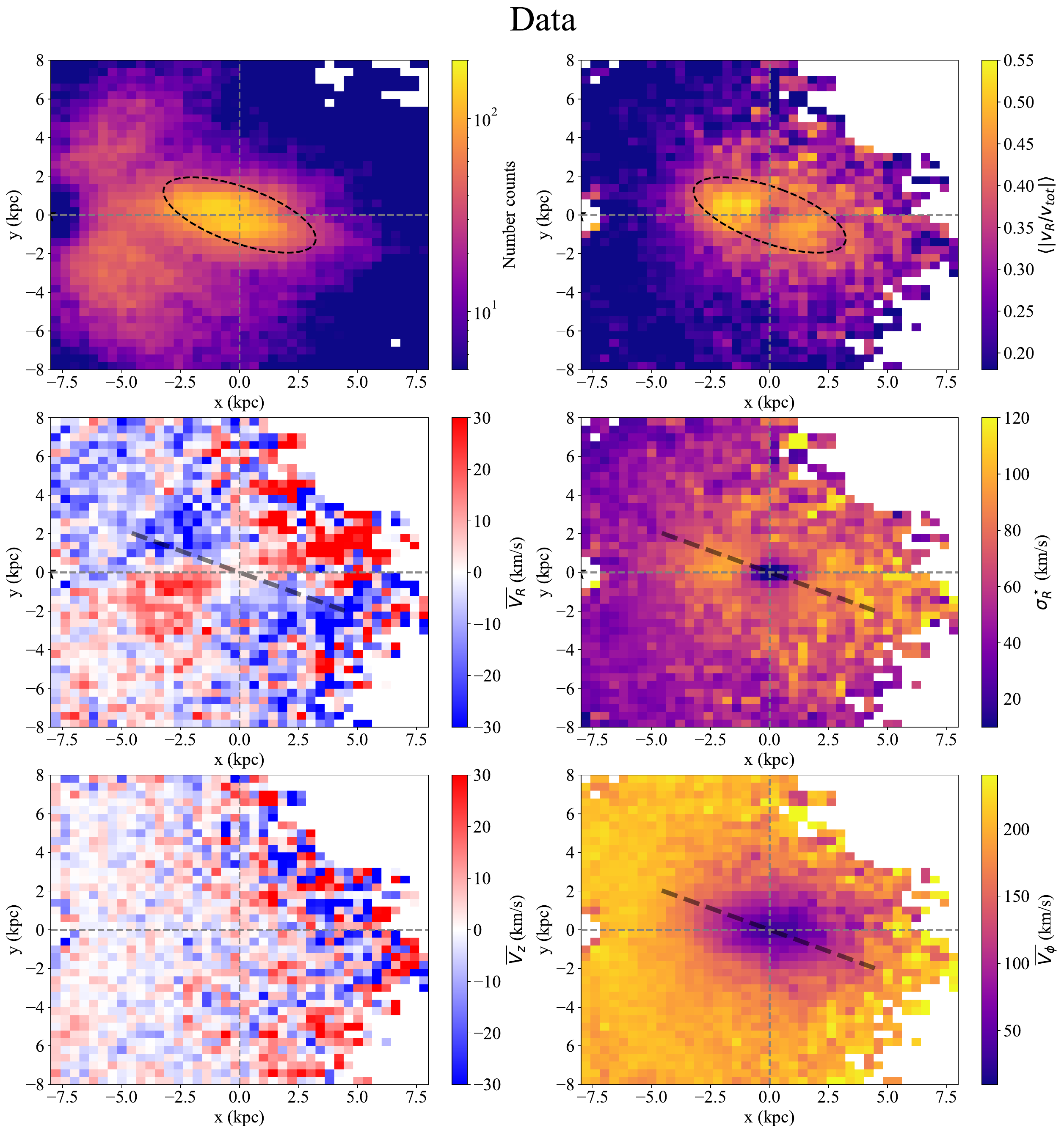}
    \caption{Spatial and kinematic maps of the LA-LPV sample. \textit{Top left:} the stellar density map. \textit{Top right:} the mean radial fraction of total velocity ($\langle |v_R/v_{\mathrm{tot}}|\rangle$) map. The black ellipse depicts the Galactic bar. \textit{Middle left:} the mean radial velocity ($\overline{V_R}$) map. The black dashed line denotes the orientation of the Galactic bar. \textit{Middle right:} the radial velocity dispersion map. \textit{Bottom left:} the mean vertical velocity ($\overline{V_z}$) map. \textit{Bottom right:} the mean azimuthal velocity ($\overline{V_\phi}$) map.}
    \label{fig::Kinematics_field_data}
\end{figure*}

We employ the approach from \citet{Gaia_Collaboration_dr3disc} by binning the stars into pixels in the $x-y$ plane. Inside each pixel, we estimate the mean velocity $\overline{V_i}$ and velocity dispersion $\sigma^\star_i$ of the velocity component $i\in(R, \phi, z)$ by deconvolving the measurement uncertainty for each star in the pixel. We write the log-likelihood as
\begin{equation}
    \ln L (\overline{V_i}, \sigma^\star_i) = -\frac{1}{2}\sum_j^N\left[ \ln(\sigma^{\star2}_i + \sigma_{i, j}^2) + \frac{(v_{i,j} - \overline{V_i})^2}{\sigma^{\star2}_i + \sigma_{i, j}^2}\right],
    \label{eqn::logL_kine}
\end{equation}
where $v_{i,j}$ and $\sigma_{i, j}$ are the $i$-component velocity and its measurement uncertainty of $j$th stars in the pixel, and $N$ is the total number of stars in the pixel. We minimise the negative log-likelihood using the Nelder-Mead method in \textsc{scipy}. We separate stars into $400\times400$~pc$^2$ squares in the $x-y$ plane and compute the mean velocity and velocity dispersion in cylindrical components $R$, $z$, and $\phi$ for each pixel. 

The radial component of the velocity is particularly interesting when studying the kinematics of the bar. We show the mean radial velocity and radial velocity dispersion in the middle row in Fig.~\ref{fig::Kinematics_field_data}. The quadrupole pattern around the Galactic centre (GC) in the mean radial velocity map is a common signature of bars seen in cosmological simulations \citep[][]{Fragkoudi_2020} and in MW observations \citep{Bovy_2019, Queiroz_2021, Leung_2023, Gaia_Collaboration_dr3disc, Liao_2024}. The pattern is caused by the streaming motion and the elongated orbits of the bar-supporting stars. The quadrupole pattern should be aligned with the major axis of the bar, but it aligns more with the line joining the Sun and the Galactic centre in our data. This effect arises because the heliocentric distance uncertainty blurs a star's observed position along the line-of-sight. This artificially elongates the bar and biases its orientation. The effect of the uncertainty is analysed using N-body simulations in the literature \citep{Vislosky_2024, Hey_2023}. We will also calibrate it later using an N-body simulation. Recently, \citet{Hey_2023} studied the kinematics of the Galactic bulge using the same tracer, LA-LPVs, in the OGLEIII-BLG survey \citep{OGLEBLG}. They show the same mean radial velocity pattern, which also extends to the other side of the disc. Moreover, the bisymmetric feature in the radial velocity dispersion with the node residing on the Galactic centre shwon in Fig.~\ref{fig::Kinematics_field_data} also agrees with the results in \citet{Gaia_Collaboration_dr3disc}.

In the bottom panels of Fig.~\ref{fig::Kinematics_field_data}, we present the mean $\overline{V_z}$, and $\overline{V_\phi}$ maps. The mean vertical velocity shows no systematic patterns on both sides of the bar, meaning that it is vertically stable at the present day. The mean azimuthal velocity drops quickly when approaching the Galactic centre, as expected from e.g. \citet{Leung_2023}. The contours of the mean azimuthal velocity map should align with the major axis of the bar in an error-free case, but the heliocentric distance uncertainty in our sample biases the contours to align more with the Sun-GC line.

We define the mean radial fraction of the total velocity as
\begin{equation}
    \Bigl\langle\left|\frac{v_R}{v_{\mathrm{tot}}}\right|\Bigr\rangle = \frac{1}{N}\sum_j^N \frac{|v_{R,j}|}{v_{\mathrm{tot},j}},
\end{equation}
where $v_{\mathrm{tot},j}$ is the total velocity of $j$th star in the pixel. The $\langle |v_R/v_{\mathrm{tot}}|\rangle$ map is shown in the top right panel of Fig.~\ref{fig::Kinematics_field_data}. As the $x_1$ orbits considered to be the backbone of the Galactic bar \citep{Patsis_2019,Petersen_2021}, are dominated by the radial motion, the region that is well-populated by $x_1$ orbits should have high $\langle |v_R/v_{\mathrm{tot}}|\rangle$. This makes the mean radial velocity fraction map a good tracer of the extent of the bar's $x_1$ orbits and hence the Galactic bar itself. As expected, $\langle |v_R/v_{\mathrm{tot}}|\rangle$ becomes higher closer to the centre, and the shape and orientation of the contours agree with a hypothetical bar that is inclined $25^\circ$ above the $x$-axis, as shown by the black dashed ellipse. Inspired by the striking, physically-motivated patterns in these kinematic projections, we conclude that the bar stars are a prominent population in our sample on both sides of the Galaxy. The full sky coverage of the {\it Gaia} sample helps to mitigate the selection effects compared to the sample in e.g. \citet{Hey_2023} and enables further quantitative analysis.

\subsection{N-body Simulation}

We present below the setups and the results of two Milky Way-analogue N-body simulations. In both cases, the Galactic bar forms within $2$~Gyr after the beginning of the simulation. We only run the N-body simulations to test the methodology and to understand the impact of distance uncertainties and the selection function (SF). Accordingly, we focus on the initialisation and basic properties of the snapshots here.

\subsubsection{Galaxy A}
The simulated galaxy has four main components: a stellar disc, a gaseous disc, a classical bulge and a halo. The stellar disc is composed of a thin and thick disc; the halo includes both dark matter and the stellar halo. An equilibrium, self-consistent model of these components is generated using the \textsc{Agama}\citep{Vasiliev_2019} \texttt{self-consistent-modelling} routine. In brief, the program provides initial conditions using action-based distribution functions in a Milky Way-like potential. Given an initial guess of the galactic potential, a density distribution is computed by calculating the zeroth-order velocity moment of the distribution functions. The distribution function is updated with the new potential from the generated density distribution, but the action-based functional form remains unchanged throughout the routine. We iterate the last two steps until they converge. A more detailed description and justification of the procedures is in \cite{Vasiliev_2019} \citep[also see][]{Tepper-Garcia_2021}. 

We employ the quasi-isothermal distribution function in Eq.~\ref{eqn::quasi-isothermal} for the disc-like structures \citep{Binney_2010} and the double-power law form in Eq.~\ref{eqn::double_power_law} that is similar to \citet{Posti_2015} and \cite{Wi15} for the spheroidal-like structures. The distribution function parameters we used to generate the initial conditions of the N-body simulation are given in Tables~\ref{tab::Nbody_discDF_params} and \ref{tab::Nbody_spheDF_params}, where the parameters listed follow the same definition as in Eq.~\ref{eqn::quasi-isothermal} and \ref{eqn::double_power_law}. Parameters not shown in the Tables are set to their default \textsc{Agama} (1.0 version) values. The initial potential is the best-fit potential from \citet{McMillan_2017}, though the choice of the initial potential only weakly affects the final self-consistent model. After the iteration as described above, in the self-consistent model, the total mass of the disc is $\sim 4.6\times10^{10}$~$M_\odot$; the halo $\sim 1.1\times10^{12}$~$M_\odot$; the bulge $\sim 9.0\times10^{9}$~$M_\odot$, which are all consistent with the value estimated for the Milky Way \citep{Bland-Hawthorn_2016}. The radial scale length of the disc, $R_d$, is $\sim2.5$~kpc. The mass of the static gaseous disc is set to $\sim 10^{10}$~$M_\odot$. 

The detailed DFs for disc and halo are given for completeness:

\begin{align}
&f(\bs{J}) = \frac{\tilde\Sigma\,\Omega}{2\pi^2\,\kappa^2} \times
\frac{\kappa}{\tilde\sigma_r^2} \exp\left(-\frac{\kappa\,J_r}{\tilde\sigma_r^2}\right) \times
\frac{\nu}   {\tilde\sigma_z^2} \exp\left(-\frac{\nu\,   J_z}{\tilde\sigma_z^2}\right) \times B(J_\phi),\nonumber\\
&B(J_\phi)=\left\{ \begin{array}{ll}  1 & \mbox{if }J_\phi\ge 0, \\
\exp\left( \frac{2\Omega\,J_\phi}{\tilde\sigma_r^2} \right) & \mbox{if }J_\phi<0, \end{array} \right.,\nonumber\\
&\tilde\Sigma(R_\mathrm{c})  \equiv \Sigma_0 \exp( -R_c / R_\mathrm{disc} ) ,\nonumber\\
&\tilde\sigma_r^2(R_\mathrm{c}) \equiv \sigma_{r,0}^2 \exp( -2R_\mathrm{c} / R_{\sigma,r} ) + \sigma_{min}^2,\nonumber\\
&
\tilde\sigma_z^2(R_c) \equiv 2h_{\mathrm{disc}}^2\nu^2(R_c) + \sigma_{min}^2,
\label{eqn::quasi-isothermal}
\end{align}

\begin{align}
&f(\bs{J}) = \frac{M}{(2\pi J_0)^3}\left[1 + \frac{J_0}{h(\bs{J})}\right]^{\Gamma}\left[1 + \frac{g(\bs{J})}{J_0}\right]^{-\beta} F(J_{\mathrm{cut}}),\nonumber \\
&F(J_{\mathrm{cut}}) \equiv \exp{\left[-\left(\frac{g(\bs{J})}{J_{\mathrm{cut}}}\right)^2\right]},\nonumber\\
&h(\bs{J}) \equiv h_r J_r + h_z J_z + (3 - h_r - h_z)\left|J_\phi\right|,\nonumber\\
&g(\bs{J}) \equiv g_r J_r + g_z J_z + (3 - g_r - g_z)\left|J_\phi\right|, 
\label{eqn::double_power_law}
\end{align}

\begin{table*}
\caption{Quasi-isothermal distribution function parameters for disc-like components. The definition of parameters follows Eq.~\ref{eqn::quasi-isothermal}.}
\begin{center}
\begin{tabular}{lcccccc}
\hline
\hline
 Components  & $\Sigma_0$ & $R_{\mathrm{disc}}$& $h_{\mathrm{disc}}$  & $\sigma_{r,0}$ & $\sigma_{\mathrm{min}}$  & $R_{\sigma,r}$ \\
  $\,$ & $(\times 10^8 M_\odot\,\mathrm{kpc}^{-2})$ & (kpc) & (kpc) & (km/s) & (km/s) & (kpc)\\
 \hline
 Thin disc & 8.9 & 2.5 & 0.2 & 100 & 5.0 & 6.0\\
 Thick disc & 1.8 & 3.0 & 0.6 & 180 & 5.0 & 6.0\\
\hline
\hline
\end{tabular}
\end{center}
\label{tab::Nbody_discDF_params}
\end{table*}

\begin{table*}
\caption{Double exponential distribution function parameters for spheroidal-like components. The definition of parameters follows Eq.~\ref{eqn::double_power_law}.}
\begin{center}
\begin{tabular}{lccccccccc}
\hline
\hline
 Components  & $M$ & $\Gamma$ & $\beta$ & $J_0$ & $h_r$  & $h_z$ & $g_r$ & $g_z$ &$J_{\mathrm{cut}}$ \\
  $\,$ & $(\times 10^9 M_\odot )$ & $\,$ & $\,$ & (kpc km/s) & $\,$ & $\,$ & $\,$ & $\,$ & (kpc km/s)\\
 \hline
 Bulge & 0.03 & 0 & 1.8 & 2.0 & 1.4 & 0.8 & 1.4 & 0.8 & 280 \\
 Stellar halo & 1.5 & 0 & 3.5 & 500 & - & - & 1.6 & 0.7 & 100,000 \\
 Dark matter halo & 4,000 & 1.2 & 3.1 & 16,000 & 1.4 & 0.8 & 1.2 & 0.9 & 20,000 \\
\hline
\hline
\end{tabular}
\end{center}
\label{tab::Nbody_spheDF_params}
\end{table*}

\begin{figure*}
    \includegraphics[width = \textwidth]{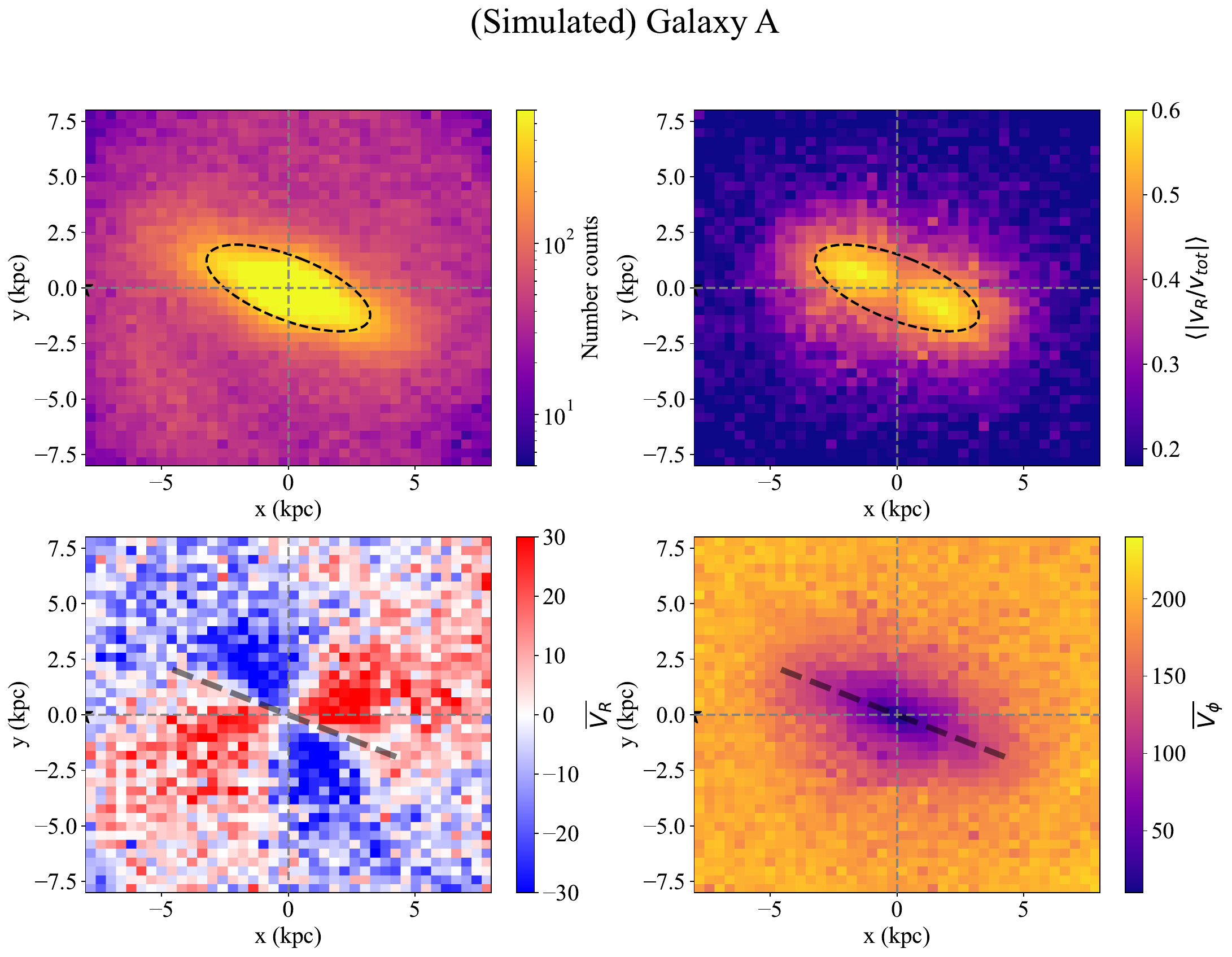}
    \caption{The stellar density (upper left), radial velocity (lower left), azimuthal velocity (lower right) and the mean radial fraction of the total velocity (upper right) map of the simulated Galaxy A. We do not show the map of $\sigma^\star_R$ and $\overline{V_z}$ as they are less interesting. The line and ellipse contour are both inclined $25^\circ$ above the x-axis to denote the orientation of the bar in this galaxy.}
    \label{fig::Kinematics_field_GalaxyA}
\end{figure*}

\begin{figure*}
    \includegraphics[width = 0.98\textwidth]{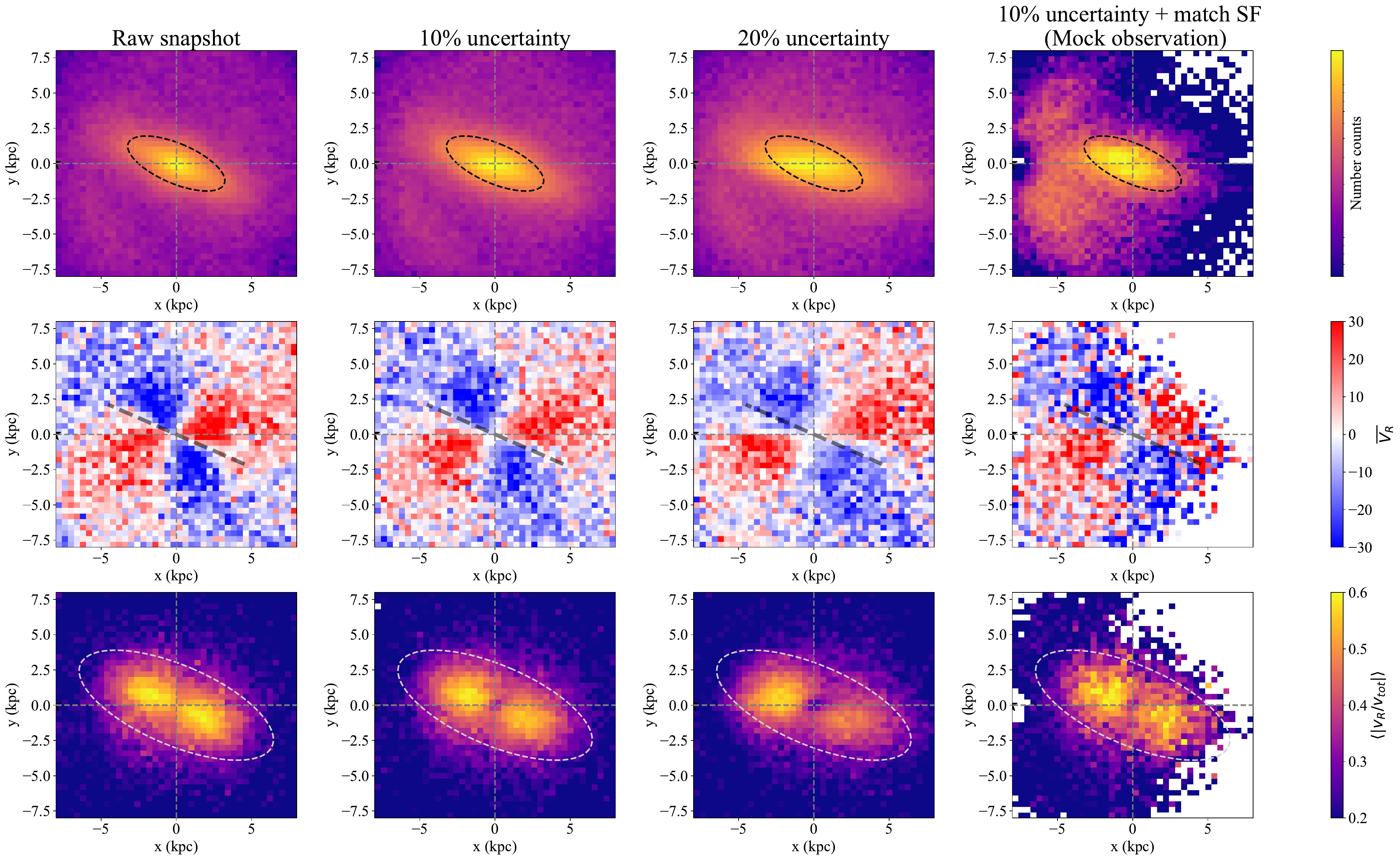}
    \caption{Effect of the heliocentric distance uncertainty and selection function on the kinematic maps of the \textit{simulated} Galaxy A. \textit{Leftmost column:} the raw stellar density and kinematic maps of Galaxy A without distance uncertainty and selection function (SF). \textit{Second left column:} the stellar density and kinematic maps of Galaxy A, in which stars are scattered with 10$\%$ distance uncertainty. \textit{Second right column:} Same but with $20\%$ distance uncertainty. \textit{Rightmost column:} the stellar density and kinematic maps of Galaxy A with $10\%$ distance uncertainty and SF matched to the observation. \textit{Top row:} the stellar density map. \textit{Middle row:} the radial velocity map. \textit{Bottom row:}  the radial fraction of total velocity map. The black dashed lines and ellipse contours in each panel are used to label the true bar angle.}
    \label{fig::Kinematics_field_GalaxyA_mockobs}
\end{figure*}

The initial conditions for the N-body simulation are sampled using the \textsc{Agama} \texttt{sampling} routine. We evolve the particles using \textsc{PyFalcon}, the Python interface of \textsc{falcON} code \citep{Dehnen_2000}, for $6$~Gyr. The softening length is set to $25$~pc, and the timestep is $0.122$~Myr. A snapshot is taken for every $0.25$~Gyr, and for every time, a consecutive snapshot is taken at $15$~Myr later to calculate the pattern speed of the bar using the finite difference method. The fraction of disc to total mass enclosed in $2.2\times R_d$ is $f_{\rm d} \sim 0.6$. This factor is a good indicator of the bar instability --in particular, the bar formation timescale is shorter for larger $f_{\rm d}$ \citep{Fujii_2018}. If $f_{\rm d} \sim 0.6$, then the bar formation timescale is well below the Hubble time. A bar formed within $1$~Gyr in our simulation, and the bar is stable and long-lived in later times. We save all the snapshots after $1$~Gyr from the start of simulation for later analysis. The pattern speeds of all the saved snapshots in simulated Galaxy A lie between $\sim 12 - 32$~km s$^{-1}$ kpc$^{-1}$. We focus on the snapshot at $2$~Gyr as a representative example of Galaxy A to analyse the bar kinematics therein. 

In the snapshot at $2$~Gyr after the beginning of the simulation, the rotation curve is $\sim 210$~km/s at the nominal Sun's location. The pattern speed, $\Omega_{\rm b}$, is $27$~km s$^{-1}$ kpc$^{-1}$ calculated with the finite difference method. The bar angle and the bar length in each snapshot are computed using the Fourier monopole and quadrupole components:
\begin{align*}
    A_m(R) = \frac{1}{\pi} \int_0^{2\pi} \Sigma(R, \phi)\cos{(m\theta)} d\theta, \\
    B_m(R) = \frac{1}{\pi} \int_0^{2\pi} \Sigma(R, \phi)\sin{(m\theta)} d\theta. \\
\end{align*}

The bar strength $A_{\rm b}$ and angle, $\phi_{\rm b}$, can be written as
\begin{equation}
     A_{\rm b}(R) \equiv \sqrt{A_2^2 + B_2^2}/A_0, \\
    \label{eqn::bar_stength}
\end{equation}
\begin{equation}
    \phi_{\rm b} = \frac{1}{2} \tan^{-1}\left(B_2/A_2\right),
    \label{eqn::bar_angle}
\end{equation}

respectively. Similar to \citet{Rosas_Guevara_2022}, we define the bar half-length $R_{\rm b}$ as the radius where $A_{\rm b}(R_{\rm b})=0.15$, and the bar half-length is then $\sim5.8$~kpc. 

\subsubsection{Galaxy B}

Further to the model described above, we add another N-body simulation, which we adopt from \citet{Tepper-Garcia_2021}. \citet{Tepper-Garcia_2021} present a three-component Milky Way-analogue galaxy that composed of a stellar bulge, a stellar disc, and a dark matter halo. We reproduce the N-body simulation in \citet{Tepper-Garcia_2021} using the same parameters. While the disc is initialised with the quasi-isothermal distribution (Eq.~\ref{eqn::quasi-isothermal}) and the corresponding parameters therein, the bulge and the dark matter halo's distribution functions are initialised directly from their defined density distribution using \texttt{QuasiSpherical} method in \textsc{Agama} that builds on the Eddington inversion formula. The resulting self-consistent model has a disc with mass $\sim 4.4 \times 10^{10}$~$M_\odot$, a bulge with $\sim 1.3 \times 10^{10}$~$M_\odot$, and a halo with mass $\sim 1.7 \times 10^{12}$~$M_\odot$. No gaseous disc is set in this simulation. 

We run the N-body simulation with the same code and parameters as in the previous model (Galaxy A). We also take snapshots with the same time step and simulate $6$~Gyr worth of evolution. In this simulation, the bar settles within $\sim 1.5$~Gyr. Hence, we save all snapshots afterwards. The pattern speeds of the bar in this simulation stays in the range between $30 - 45$~km s$^{-1}$ kpc$^{-1}$, measured using the finite difference method. We choose the snapshot at $t=4$~Gyr to represent Galaxy B. The pattern speed of the bar in this snapshot is $32.56$~km s$^{-1}$ kpc$^{-1}$, similar to that of the Milky Way (see the discussion above).. 

\subsection{Kinematics of Galaxy A}

In a similar way to the handling of the {\it Gaia} data, we plot the kinematics of Galaxy A by binning stars in pixels in the $x-y$ plane as shown in Fig.~\ref{fig::Kinematics_field_GalaxyA}. We rotate Galaxy A so that the major axis of the bar also inclines $25^\circ$ ahead of the $x$-axis to mimic the setup with an observer at the Sun's location in the actual Milky Way. A clear quadrupole pattern is visible in the mean radial velocity map of Galaxy A as shown in the lower left panel in Fig.~\ref{fig::Kinematics_field_GalaxyA}. The same quadrupole pattern is seen in other simulated barred galaxies. As shown in the top right panel in Fig.~\ref{fig::Kinematics_field_GalaxyA}, the value of $\langle |v_R/v_{\mathrm{tot}}|\rangle$ is high in the bar region as expected due to the orbital properties of the bar-supporting stars.

To fairly compare the simulation with observations, we repeat the analysis in \citet{Hey_2023} and \citet{Vislosky_2024}. We scatter the particles in Galaxy A with heliocentric distance uncertainties and investigate how this propagates to the kinematics. We neglect the proper motion and radial velocity uncertainty in this analysis as they are less significant than the distance uncertainty. The top row of Fig.~\ref{fig::Kinematics_field_GalaxyA_mockobs} shows the stellar density map of Galaxy A with different heliocentric distance uncertainties. The heliocentric distance uncertainties blur the shape of the bar as expected. By smearing the position of the stars along the line-of-sight direction, the bar is artificially stretched and biased towards the Sun-GC line. In the middle panel of Fig.~\ref{fig::Kinematics_field_GalaxyA_mockobs}, we present the mean radial velocity field of Galaxy A for different fractional heliocentric distance uncertainties. The black dashed line always shows the orientation of the true major axis of the bar. In an error-free setup, the quadrupole pattern changes signs when crossing the major axis of the bar. However, in the presence of $20\%$ heliocentric distance uncertainty, the sign-switching position changes from the true major axis of the bar to the Sun-GC line. We also find the $\langle |v_R/v_{\mathrm{tot}}|\rangle$ map is less affected by the heliocentric distance uncertainty as shown in the bottom panels. While the $\overline{V_R}$ map has sign-switching along with the Sun-GC line, the orientation of the $\langle |v_R/v_{\mathrm{tot}}|\rangle$ map is less biased and preserves better the original angle of the bar as shown by the white contour. This happens because the distance uncertainty affects both the numerator and the denominator in $\langle |v_R/v_{\mathrm{tot}}|\rangle$ ratio, and thus its impact appears to approximately cancel out. Hence, we argue that we can use the signal in the measured $\langle |v_R/v_{\mathrm{tot}}|\rangle$ map to estimate the angle of the bar.


We further apply the observational selection function to Galaxy A to mimic the properties of the {\it Gaia} sample. The selection function is implemented by identifying and retaining the mock particle (in Cartesian coordinates) closest to an observed star. We discard simulated particles that either fail to become a closest match in $xyz$ or for which the distance between the real star and the closest-matched mock particle is greater than $0.1$~kpc. After matching the data and simulation this way, the simulated stars carry the effects of the selection function and have roughly the same number counts as the dataset while also mimicking the spatial distribution. In the rightmost column of Fig.~\ref{fig::Kinematics_field_GalaxyA_mockobs}, we show the kinematics of Galaxy A after scattering particles by $10\%$ distance uncertainty and applying the selection function as described above. In the top right panel, the stellar density map, after applying the selection function, is identical to that in Fig.~\ref{fig::Kinematics_field_data} by design. We highlight the similarities compared to the kinematics of our dataset in Fig.~\ref{fig::Kinematics_field_data}. Testing the robustness of these kinematic maps by adding $\sim 5\%$ systematic distance bias, we conclude that the maps are qualitatively the same with a minor distance bias. This reinforces the validity of our sample uncertainties and analysis while also demonstrating the power of this sample in studying the Galactic bar.

\subsection{A kinematic measurement of the Galactic bar length}
\label{sec::barlength}

\begin{figure*}
    \includegraphics[width = 0.99\textwidth]{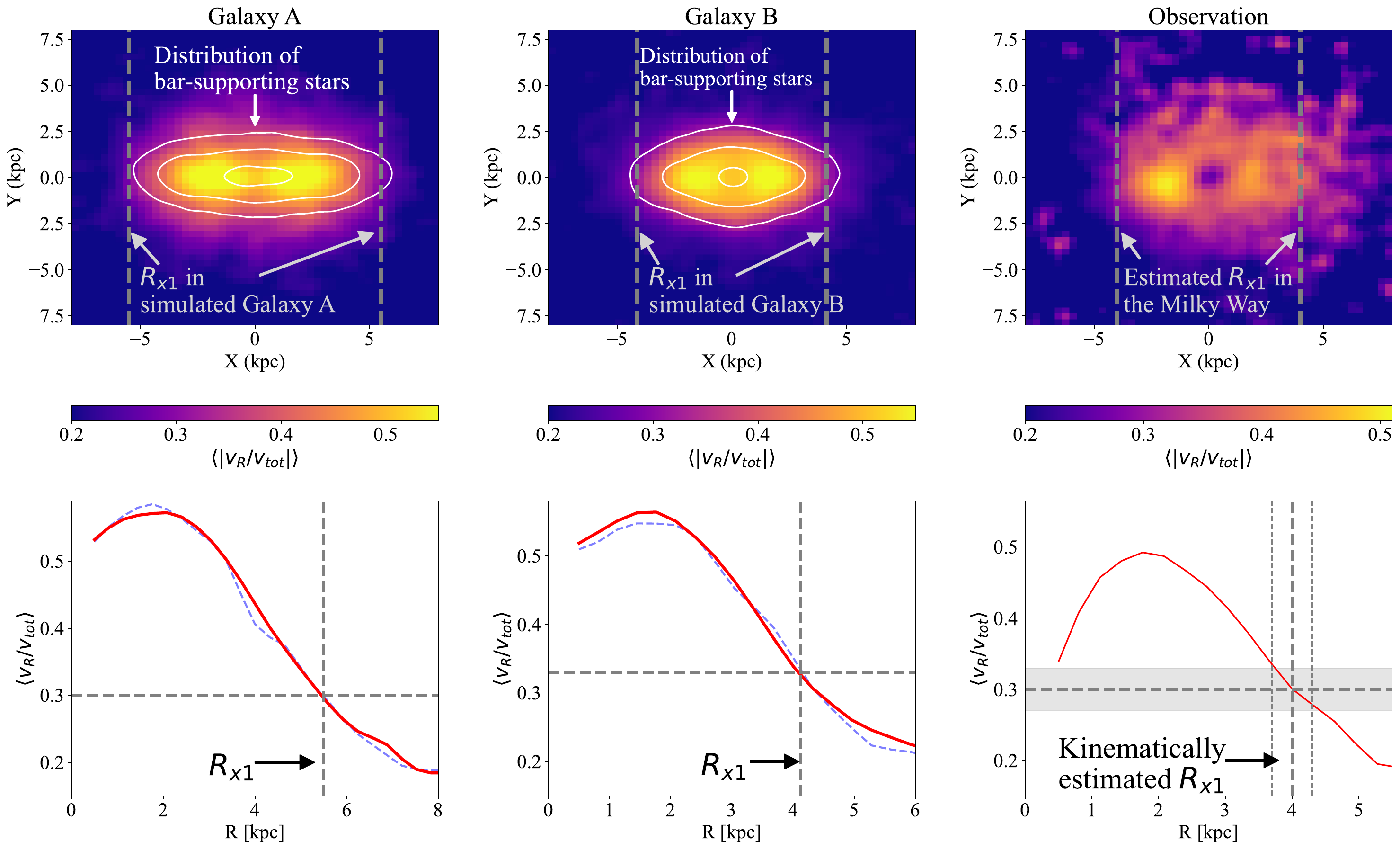}
    \caption{The comparison between the signal in the $\langle |v_R/v_{\mathrm{tot}}|\rangle$ map and the spatial distribution of the bar-supporting stars. \textit{Top row:} the $\langle |v_R/v_{\mathrm{tot}}|\rangle$ map in Galaxy A (left column), B (middle column) and observed Milky Way (right column). Observed stars in the Milky Way are rotated $25^\circ$ anti-clockwise to align the major axis of the bar with the X-axis. The white contours in the \textit{top left} and \textit{top middle} panel show the distribution of bar-supporting stars in Galaxy A and B selected by the orbital frequency ratios. The grey dashed lines denote the dynamical length of the galactic bar, $R_{x_1}$, in Galaxy A and B. \textit{Bottom row:} the radial profile of the $\langle |v_R/v_{\mathrm{tot}}|\rangle$ value along the bar major-axis on the side that close to the Sun. The red solid lines and blue dashed lines are the radial $\langle |v_R/v_{\mathrm{tot}}|\rangle$ profile before and after applying the observational effects (distance uncertainty + SF). The vertical grey dashed line in the left and middle panel labels the dynamical length of the bar in Galaxy A and B, and they intersect with the radial $\langle |v_R/v_{\mathrm{tot}}|\rangle$ profile at $\langle |v_R/v_{\mathrm{tot}}|\rangle\approx0.3$. \textit{Bottom right:} the radial profile of $\langle |v_R/v_{\mathrm{tot}}|\rangle$ of the Milky Way. The horizontal grey band denotes the range for which $\langle |v_R/v_{\mathrm{tot}}|\rangle\approx0.3$. It intersects with the red line at $R\approx4.0$~kpc denoted by the vertical dashed lines, which gives a kinematic estimation of the dynamical length of the Milky Way.}
    \label{fig::bar_length_estimate}
\end{figure*}

The length of the Galactic bar is usually defined through the stellar overdensity or through analysis of the apocentre distribution of bar-supporting stars \citep{Athanassoula_Misiriotis_2002, Petersen_2021, Lucey_2023}. \citet{Athanassoula_Misiriotis_2002} proposed finding the bar length by fitting ellipses to the stellar density. The Fourier quadrupole mode is also frequently used to define the bar length. The radius at which the bar strength, $A_b$ in Eq.~\ref{eqn::bar_stength}, falls below some threshold ($20\%$ of the maximum value in $A_b$ is adopted for the bar length in \citealt{Athanassoula_Misiriotis_2002}). The bar length varies when different thresholds are chosen \citep{Athanassoula_Misiriotis_2002, Rosas_Guevara_2022}. Also, \citet{Hilmi_2020} showed that the bar length estimates using these methods are biased when a spiral arm connects to the bar. The dynamical length defined by the extent of the bar-supporting orbits is less affected by the connected spiral arm \citep{Petersen_2024}. 

Detecting bar-supporting orbits is usually done either by selecting stars through orbital frequency ratios \citep{Portail_2015, Lucey_2023} or through the compactness of the orbital apocentre in a few orbital timescale to find the $x_1$ orbits \citep{Petersen_2021}. In either case, the classification requires the knowledge of the gravitational potential in the inner Galaxy, as we need to integrate orbits to study the orbital families. However, as the signature of bar orbits is likely the cause of the high-value signal in the $\langle |v_R/v_{\mathrm{tot}}|\rangle$ map, we explore the possibility of using the $\langle |v_R/v_{\mathrm{tot}}|\rangle$ map to define the dynamical length of the bar. This is similar to the method in \citet{Petersen_2024}, which uses the mean $v_{\perp}$ map to detect the oscillation in the $x_1$ orbits, where $v_{\perp}$ is the velocity in the direction perpendicular to the bar major axis. \citet{Petersen_2024} used the ratio of the octupole and quadrupole moment of the mean $v_{\perp}$ map to track the extent of the $x_1$ orbits. However, as the method involves knowledge of higher-order Fourier components, it is necessary to have full coverage and high signal-to-noise ratio measurements in the inner Galaxy. Therefore, the method is more suitable for simulations and extragalactic observations than for applications to the Milky Way. 

To test the hypothesis that the signal in $\langle |v_R/v_{\mathrm{tot}}|\rangle$ is linked of the exact mix bar orbits, we turn to the simulations. The $\langle |v_R/v_{\mathrm{tot}}|\rangle$ map of Galaxy A and B are presented in the top left and middle panel in Fig.~\ref{fig::bar_length_estimate}. To classify the orbits, we use the multipole expansion method implemented in \textsc{Agama} to obtain the potential of the snapshot, and then integrate the particles for $2$~Gyr. The pattern speed of the potential is again calculated using the finite difference method. For each particle, we compute the Fast Fourier Transform (FFT) of the time series in $R$ and $X$ coordinates and find the primary peaks as the orbital frequencies, where $X$ is aligned with the major axis of the bar. We denote $\Omega_i$ as the orbital frequency in the $i$-axis. The distinction between a bar and a disc particle is simple in practice. We use a cut similar to that in \citet{Portail_2015} and assign stars with $\Omega_R/\Omega_x\in 2\pm 0.2$ and $r_{\rm apo}<7$~kpc to the bar, where $r_{\rm apo}$ is the spherical apocentric radius of the orbits. Although cutting on the apocentric radius appears to conflict with the purpose of this test, we note that the resulting extent of bar orbits is far smaller than $7$~kpc. We only use this cut to remove the disc contamination that has an orbital frequency ratio satisfying the bar selection by coincidence. The contours of the distribution of selected bar stars are shown in the top and middle panels of Fig.~\ref{fig::bar_length_estimate} in white. The distribution of the bar stars matches beautifully to the shape of the signal in the $\langle |v_R/v_{\mathrm{tot}}|\rangle$ map, reinforcing the idea that it can be used as a kinematic tracer of bar orbits.

Additionally, amongst these selected bar stars, we identify $x_1$ orbits, namely, we apply the method proposed in \citet{Petersen_2021}. As a short summary of the method, the $x_1$ orbits are selected by examining the spatial compactness of the apocentre distribution at the two ends of the bar for each trajectory. The 2D spatial information of $20$ consecutive apocentres is recorded for each trajectory, and the k-mean clustering method \citep[][$k=2$ corresponding to two ends of the bar]{Lloyd_1982} is employed to assess the compactness of these apocentres relative to the ends of the bar. The expected purity of $x_1$ orbits selected using this method is $\sim99\%$ \citep{Petersen_2021}, but we also visually inspected all of the selected $x_1$ orbits to ensure the algorithm performs as expected. The bar lengths estimated from the radius spanned by $x_1$ orbits, denoted as $R_{x_1}$, are $5.5$ and $4.1$~kpc for Galaxies A and B and are labelled by the vertical grey dashed line in the corresponding panel in Fig.~\ref{fig::bar_length_estimate}. $R_{x_1}$ agrees with the outskirts of $\langle |v_R/v_{\mathrm{tot}}|\rangle$ signal in both simulated galaxies. In the bottom panels of Fig.~\ref{fig::bar_length_estimate}, we show the $\langle |v_R/v_{\mathrm{tot}}|\rangle$ value as a function of radius along the major axis of the bar in red solid lines. The vertical dashed line again labels $R_{x_1}$, and in both galaxies corresponds to $\langle v_R/v_{\mathrm{tot}}\rangle \approx 0.3$. Hence, we use $\langle v_R/v_{\mathrm{tot}}\rangle|_{R_{\rm b}} = 0.3$ as a kinematic estimation of the bar length that is calibrated to agree with $R_{x_1}$. The method is also tested and proved to be robust with the distance uncertainties ($10\%$) and the selection functions. The radial $\langle |v_R/v_{\mathrm{tot}}|\rangle$ profile along the bar major axis after imitating the observation is shown by the blue dashed line in Fig.~\ref{fig::bar_length_estimate}, in which the difference between the red and blue lines are negligible.

Applying this kinematic bar length measurement method to the Milky Way, we first rotate the stars in our sample by $25^\circ$ to align the $X$-axis with the bar major axis. We fill the $(X,Y)$ pixels that have no stars with $\langle |v_R/v_{\mathrm{tot}}|\rangle$ equals $0$. We plot the $\langle |v_R/v_{\mathrm{tot}}|\rangle$ map of the rotated Milky Way in the top right panel of Fig.~\ref{fig::bar_length_estimate}, and in the bottom, we plot the $\langle |v_R/v_{\mathrm{tot}}|\rangle$ value as a function of radius along the bar major axis on the side facing the Sun. The $\langle v_R/v_{\mathrm{tot}}\rangle=0.3$ corresponds to $R_{\rm b}\approx4.0$~kpc. Hence, we report $R_{\mathrm{b, kine}}\approx4.0$~kpc as a kinematic analogue of the dynamical length of the Milky Way bar. We denote this length by $R_{\mathrm{b,kine}}$ to emphasise that this is not a dynamical measurement but is instead a kinematic analogue. Note that this measurement of the bar length is independent of the Galactic potential and does not require the completeness of the sample as long as the sample gives enough coverage and sampling of the inner Galaxy. 

\subsection{Pattern speed estimates from the continuity equation}
\label{sec::patternspeed_measurement}
We can measure the pattern speed of the Galactic bar using the kinematics of LA-LPVs in our sample with the aid of the continuity equation. A critical issue when measuring the pattern speed of the Galactic bar using the continuity equation is the incompleteness of the observations as the stellar density is involved in the calculation. \citet{Bovy_2019} find the pattern speed of the Galactic bar by applying the continuity equation to APOGEE stars with spectroscopic \texttt{AstroNN} distances. In \citet{Bovy_2019}, the selection function issue is avoided by postulating that the bar density is stratified on concentric ellipses. \citet{Dehnen_2023} provides a method of estimating the pattern speed of the bar in a single simulation snapshot based on the continuity equation. They prove the method recovers the pattern speed of simulated galaxies, which has perfect completeness. The method in \citet{Dehnen_2023} is also different to that in \citet{Bovy_2019}, in which \citet{Dehnen_2023} compute the pattern speed by integrating the disc plane modulo a weighting function, while \citet{Bovy_2019} bin stars into radial annuli. The weighting function proposed in \citet{Dehnen_2023} behaves similarly to the selection function. Hence, choosing a proper weighting function also helps to mitigate the completeness problem. However, the weighting function in \citet{Dehnen_2023} is only a function of the cylindrical radius (see Eq. 25 in \citealt{Dehnen_2023}), so the selection function in the azimuthal direction still biases the pattern speed estimates. 

\begin{figure}
    \centering
    \includegraphics[width = \columnwidth]{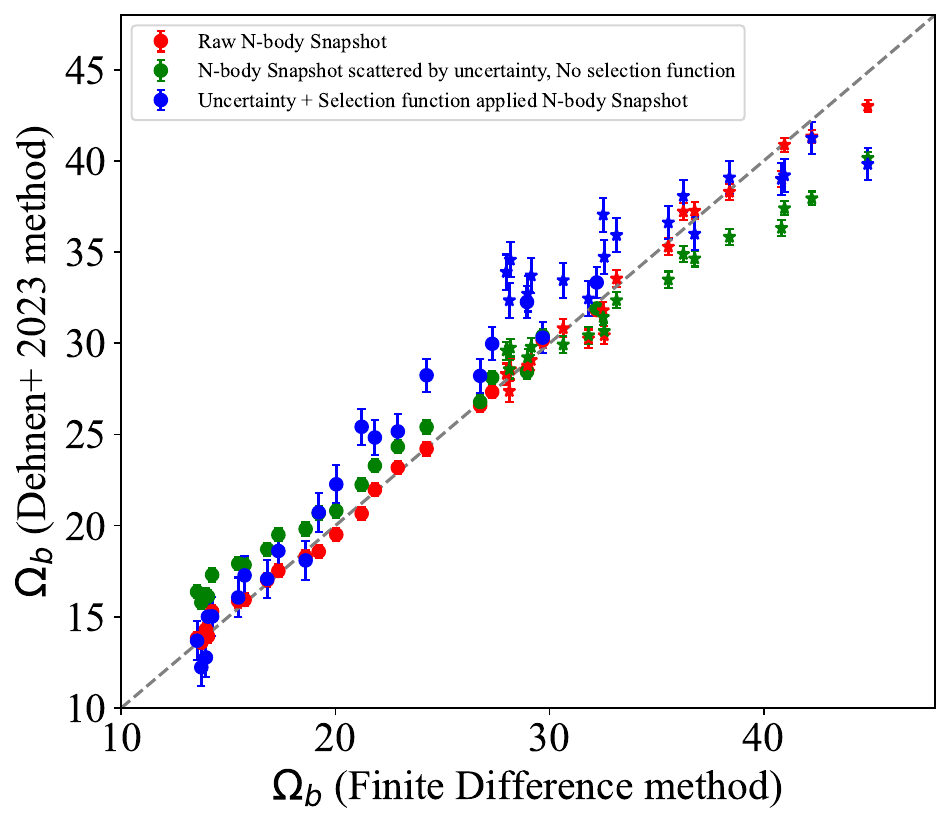}
    \caption{Validation of the pattern speed recovery using the method in \citet{Dehnen_2023} with the presence of observational caveats. The ground truth pattern speed in each N-body simulation snapshot is on the $x$-axis and the calculated pattern speed is on the $y$-axis. The circle dots are test results from N-body snapshots in Galaxy A, and the star dots are those in Galaxy B. The red dots show the pattern speed recovered when applying the method to the raw N-body snapshots without observation effects. The green dots show the pattern speed recovered when $10\%$ distance uncertainty is applied to the snapshots, and the blue dots show the pattern speed recovered when $10\%$ distance uncertainty is applied and SF is matched to the observation. The grey dashed line shows the 1:1 line to the ground truth pattern speed.}
    
    \label{fig::test_dehnen2023_method}
\end{figure}

As seen in the stellar density map shown in the top-left panel in Fig.~\ref{fig::Kinematics_field_data}, the density distribution tracks the shape of the bar and is roughly symmetric in the inner few kpc of the Galaxy. Compared to the APOGEE \texttt{AstroNN} sample in \citet{Bovy_2019, Leung_2023}, we suffer less from the completeness issue in the $x-y$ plane at the inner Galaxy. This is because of the deep photometry from {\it Gaia} and the pure photometric selection of the LA-LPV candidates in our sample. The extinction does not seriously affect the construction of the catalogue, except for the missing $\sim 150$ pc above and below the plane. Hence, we attempt to directly apply the Fourier method in \citet{Dehnen_2023} to our sample with a weighting function of $R_0 = 0.1$~kpc, $R_1 = 6$~kpc, and $R_m = 3$~kpc, where these parameters are defined by Eq.25 in \citet{Dehnen_2023}. To understand the magnitude of bias due to the effects of heliocentric distance uncertainty and the selection function, we perform a test using the two N-body simulations presented above. We implement this method to the saved N-body snapshots with manually added uncertainties and selection functions for the two sets of N-body simulations (the same procedure as we used in the final column of Fig.~\ref{fig::Kinematics_field_GalaxyA_mockobs}). Treating the pattern speed calculated from the finite difference method as the ground truth, we first apply the methods on the raw N-body snapshots ($0$ uncertainty, no incompleteness) to verify the method is implemented correctly. Then, we scatter the particles in the N-body simulation with $10\%$ heliocentric distance uncertainty, and we further mimic the spatial selection function (SF) of our sample by matching the stars in the simulated galaxies to the observed spatial distribution. The method is tested on all three scenarios (1. no error, no SF; 2. $10\%$ error, no SF; 3. $10\%$ error, SF matched to observation) as validation. The result of this test is presented in Fig.~\ref{fig::test_dehnen2023_method}. Although both heliocentric distance uncertainty and incompleteness introduce bias in the pattern speed estimation, the test results generally agree with the ground truth pattern speed because neither effect is strong in our sample. Our test galaxies cover a range in pattern speeds and bar lengths and come from two independent N-body simulations. Hence, we argue that the Fourier method in \citet{Dehnen_2023} can be directly applied to our sample and give a faithful estimation of the bar pattern speed.

Applying the above method to our sample gives the pattern speed of $34.1 \pm 0.6$~km s$^{-1}$ kpc$^{-1}$. The uncertainty here is statistical and is propagated using the Monte Carlo method, in which we repeat 1000 random realisations of the measurements with the associated uncertainties. We calculate the pattern speed for each realisation and report the mean and standard deviation. With the N-body simulation test, we can quantify the systematic uncertainty from the incompleteness and uncertainty effects on the continuity equation by calculating the standard deviation of the pattern speed residual. The systematic uncertainties brought by the selection effects are $\sim2.3$~km s$^{-1}$ kpc$^{-1}$. 

As applying the continuity equation to a sample with uncertainty and incompleteness can introduce systematic biases, we suggest that the pattern speed reported above is an estimate instead of a robust measurement. 
The pattern speed is consistent with other measurements when no assumption on the bar density distribution is made. To precisely get the pattern speed, more complicated methods should be employed, and we leave this for future works. 

\section{Orbital families in the Galactic bar}
\label{sec::dynamics}
\label{sec::orbital_family}

To analyse the dynamics of the Galactic bar, we need to assume a gravitational potential to integrate the orbits of stars. We employ the bar potential in \citet{Sormani_2022}, which is an analytic approximation of the M2M barred Milky Way model in \citet{Portail_2017}. Hence, all the results in this section are model-dependent. The chosen pattern speed for the potential is $34.1$~km s$^{-1}$ kpc$^{-1}$, motivated by our measurements in Section~\ref{sec::patternspeed_measurement}, and we integrate stars for $1$~Gyr. We use the same procedure in Section~\ref{sec::barlength} to select bar stars and stars on $x_1$ orbits. Cutting on orbital frequency ($1.8<\Omega_R/\Omega_X<2.2$) and removing stars with apocentre greater than $6$~kpc, we select a subsample of 1997 bar-supporting stars. Among these selected bar stars, $\sim 90\%$ have $\Omega_X/\Omega_Y\approx1$, which are $x_1$ orbit candidates, while the remaining $\sim10\%$ have $\Omega_X/\Omega_Y$ between $0.6$ and $0.8$. Using the rigorous method of selecting $x_1$ orbits proposed in \citet{Petersen_2021}, we further confirm $625$ stars belong to $x_1$ orbital families. The spatial extent of the $x_1$ orbits is $R_{x_1}\sim 3.8$~kpc, which is consistent with our kinematic analogue of the dynamical length measured in Section~\ref{sec::barlength}. 

With a visual inspection of stars that failed the $x_1$ orbit selection using the method in \citet{Petersen_2021}, we find the majority still belong to the $x_1$ family tree. Most of them are bifurcated $x_1$ orbits with more loosely spaced apocentres; only a small fraction of the bar-supporting stars are not members of the $x_1$ orbital family. This confirms that the $x_1$ orbital family are still the main building block of the Galactic bar \citep{Patsis_2019, Petersen_2021}.

\begin{figure}
    \centering
    \includegraphics[width = \columnwidth]{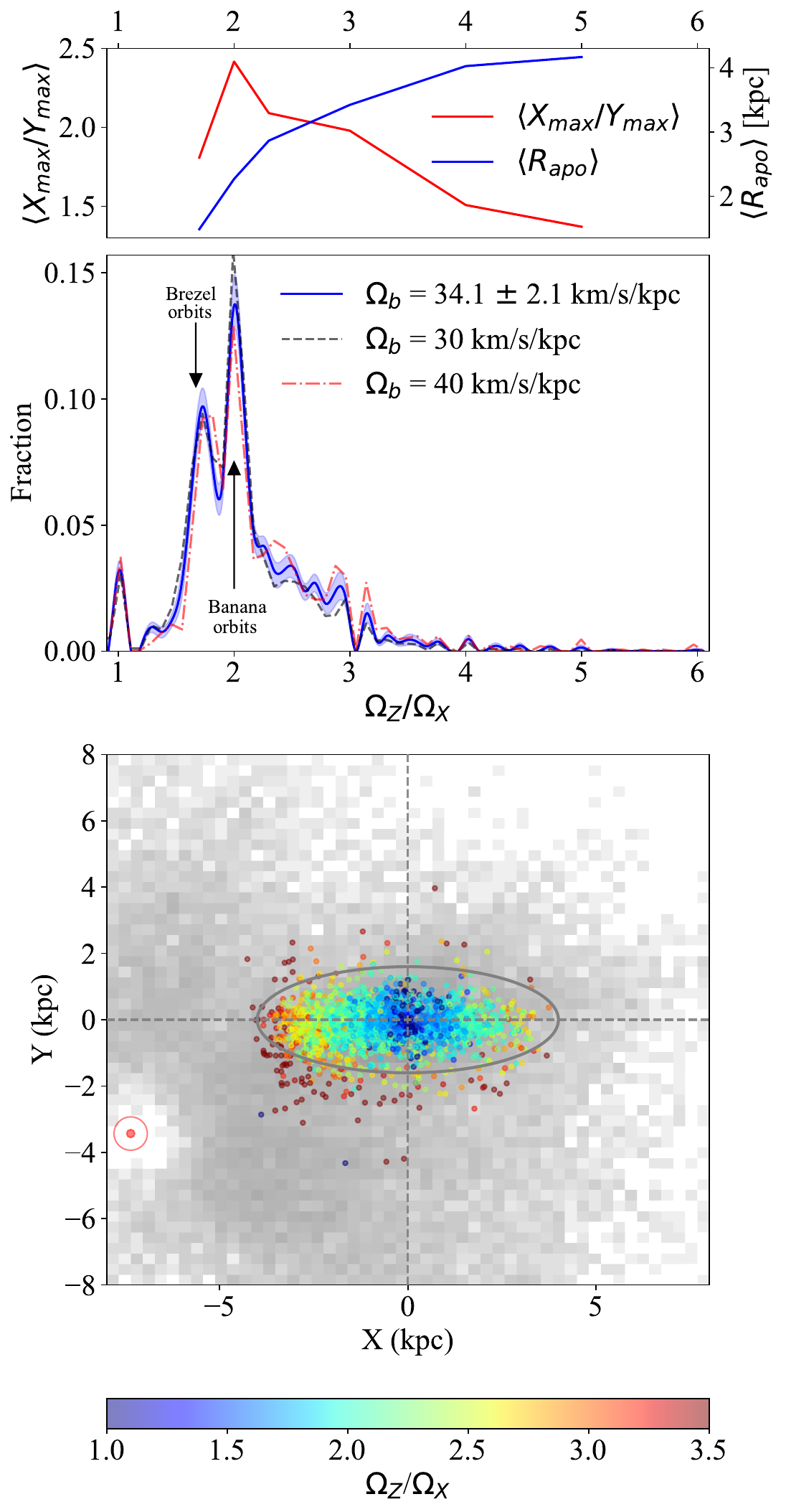}
    \caption{The orbital families in the Milky Way bar-supporting stars in terms of $\Omega_Z/\Omega_X$ values. \textit{Top:} the mean apocentre, $\langle R_{\mathrm{apo}} \rangle$ (blue line), and mean $\langle X_{\mathrm{max}}/Y_{\mathrm{max}} \rangle$ (red line) for stars as a function of the $\Omega_Z/\Omega_X$ values. \textit{Middle:} the distribution of $\Omega_Z/\Omega_X$ for bar-supporting stars in the Milky Way. The blue line and the blue-shaded region are the $\Omega_Z/\Omega_X$ distribution and the associated $1\sigma$ uncertainty when the pattern speed is $34.1\pm2.1$~km~s$^{-1}$kpc$^{-1}$. The grey and red dashed lines are the $\Omega_Z/\Omega_X$ when the pattern speed is $30$ and $40$~km~s$^{-1}$kpc$^{-1}$. The arrow labels the $\Omega_Z/\Omega_X$ values of the brezel and banana orbits. \textit{Bottom:} the spatial distribution of bar-supporting stars is shown in the foreground dots, for which each star is coloured according to its $\Omega_Z/\Omega_X$ value. Bar-supporting stars with larger $\Omega_Z/\Omega_X$ values reached farther distances from the Galactic centre. The background 2D histogram shows the distribution of LA-LPV candidates in our sample rotated $25^\circ$ to align the bar  major-axis to the $X$-axis.}
    
    \label{fig::orbital_family_data}
\end{figure}

We can study the orbital families of bar-supporting stars using the ratio $\Omega_Z/\Omega_X$, where $\Omega_Z/\Omega_X$ is the ratio between the orbital frequencies in the vertical and bar major-axis direction. Stars with different $\Omega_Z/\Omega_X$ values have different edge-on orbital trajectories \citep{Portail_2015}. We plot the $\Omega_Z/\Omega_X$ distribution of bar stars in the middle panel of Fig.~\ref{fig::orbital_family_data}. The blue line shows the orbital family distribution in our selected bar sample. The error is propagated using the Monte Carlo method, in which the pattern speed of the bar potential also participates in the random realisation. The effects of both the observational and pattern speed uncertainty are shown by the blue-shaded region in Fig~\ref{fig::orbital_family_data} to represent the $1\sigma$ level scatter. We also perform the same analysis using different pattern speeds and find only a minor effect for the orbital families (as shown in the dashed red and black lines). An obvious trend in the spatial extension and orbital shape as a function of the $\Omega_Z/\Omega_X$ is seen. We bin the bar-supporting stars in $\Omega_Z/\Omega_X$ segments and calculate the mean apocentre radius, $\langle R_{\mathrm{apo}} \rangle$, and $\langle X_{\mathrm{max}}/Y_{\mathrm{max}} \rangle$ for stars in the bin, in which $X_{\mathrm{max}}$ and $Y_{\mathrm{max}}$ are the maximum extent of an orbit along the bar major and minor axis. As shown in the top panel, orbits with greater $\Omega_Z/\Omega_X$ extend farther from the Galactic centre and become more circular. Hence, with these results, we suggest no cuts on $\Omega_Z/\Omega_X$ should be made when measuring the length of the bar because doing so could artificially shorten the measured bar length. We further illustrate this by plotting the spatial $X-Y$ distribution of the bar-supporting stars and colouring them by $\Omega_Z/\Omega_X$ in the bottom panel of the Figure. The stars with higher $\Omega_Z/\Omega_X$ reside in the outer region, which is similar to the simulated galaxies and M2M models in \citet{Portail_2015}. The grey ellipse has a semi-major axis of $4$~kpc, which we adopt from our $R_{b,\mathrm{kine}}$ measurement in section~\ref{sec::barlength}. This plot demonstrates a reassuring consistency between our kinematic measurement of the bar length and the extent of the dynamically identified of bar-supporting stars.

\begin{figure}
    \centering
    \includegraphics[width = \columnwidth]{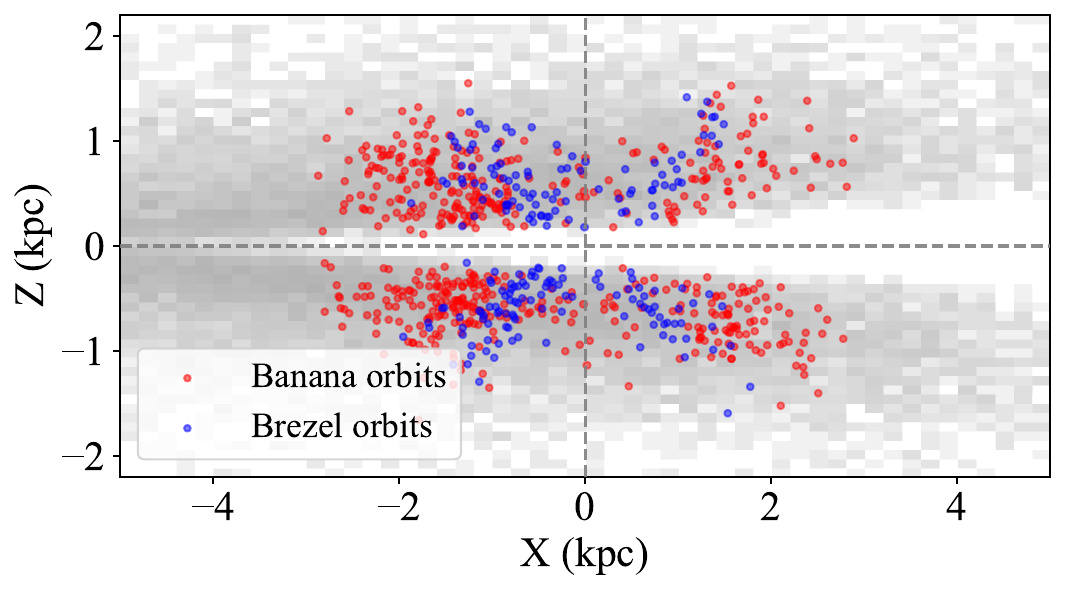}
    \caption{The spatial distribution of the stars that are assigned as on banana orbits (red dots) and brezel orbits (blue dots). There are 559 stars ($\sim26\%$ of the total bar stars) on banana orbits, and 209 stars ($\sim10\%$ of the total bar stars) on brezel orbits. The background is the 2D histogram of all stars in our sample to show the selection function of our sample.}
    \label{fig::X_shape_family}
\end{figure}

Compared to the M2M model of the Milky Way in \citet{Portail_2015}, we find a wider distribution in $\Omega_Z/\Omega_X$. Most stars in our sample have $\Omega_Z/\Omega_X<3$ and there is a significant sub-population of stars with $\Omega_Z/\Omega_X\approx2$. Orbits with $\Omega_Z/\Omega_X=2$ are particularly interesting as they belong to $x_1 v_1$ family, sometimes called "banana" orbits due to their particular side-on projection~\citep[e.g.,][]{Sk02, Wi16}. These stars exhibit a $2:1$ resonance in the vertical direction and used to be considered as the backbone of the X-shape in the boxy/peanut bulge due to the shape of their orbits. However, M2M models of the Milky Way bulge in \citet{Portail_2015} also revealed that "brezel" orbits can become a dominant contributor to the Galactic X-shape. The brezel orbits are stars in $5:3$ vertical resonances ($\Omega_Z/\Omega_X\approx5/3$). Unlike the orbital families distribution in \citet{Portail_2015}, we still find banana orbits dominating with $\sim15\%$ fractional contribution to the overall population. We present the spatial $X-Z$ distribution of stars belonging to banana orbits and brezel orbits in Fig.~\ref{fig::X_shape_family}. Both orbital families trace the X-shaped bar; banana orbits contribute more stars, while the brezel orbits show a sharper X-shape. We also present the side-on projection of each orbital family in Appendix~\ref{Appendix::side_on_projection} to illustrate their contribution to bar density. Intriguingly, as revealed in Fig.~\ref{fig::X_shape_family}, the X-shape structures contributed by different orbital families are different due to their spatial extension. We suggest that this signature provides scope for determining the more detailed structure in the X-shape, and also the main orbital family in the Galactic X-shaped structure.

We also test the impact of the selection function by applying the previously used spatial matching technique on the Galaxy A. We test the selection function effect only on Galaxy A because it has similar orbital families to the Milky Way (see Fig.~\ref{fig::appendix::SF_orbits} in Appendix~\ref{Appendix::SF_on_orbits}). We find that the selection function in our sample only gives a small adjustment from the raw orbital family distribution (see Fig.~\ref{fig::appendix::SF_orbits}). The main selection effects we have are 1) stars in $\sim150$ pc above and below the plane are missing 2) the outskirts of the bar on the far side of the disc are less complete and suffer greater distance uncertainty. After applying the selection function, the relative fraction of stars with banana orbits is slightly enhanced as they spend more time outside the Galactic plane. Stars with $\Omega_Z/\Omega_X\gtrsim2.5$ are slightly depleted because we miss some of these stars on the far side of the Galaxy. A more detailed discussion of the selection function impact is in Appendix~\ref{Appendix::SF_on_orbits}. However, due to the satisfactory completeness of this sample in the inner Galaxy, both effects are mild. Therefore, we conclude that the banana orbits, $x_1 v_1$ family, contribute most to the Galactic X-shape bar, but contributions from higher-order resonances are also non-negligible.

\section{Discussion and Comparison}
\label{sec::discussion}
\citet{Hilmi_2020} show that the bar-spiral interaction could bias the instantaneous measurements of the Galactic bar parameters. The bar length can be overestimated due to the bar-spiral coupling and constructive interference in the overlapping region. They also argued that the pattern speed measurement using the modified Tremaine-Weinberg method in \citet{Sanders_2019} can lead to a fluctuation of $10-20\%$ around the mean when the bar is connected to a spiral arm. The Scutum–Centaurus arm is likely connected to the spiral arm at the present day \citep{Rezaei_Kh_2018}. We devote this section to discussing the implications of the bar-spiral interaction in our results and comparing our measurements with those of previous studies.  

\subsection{Bar length}

We do not measure the bar length using any density-based method due to the bar-spiral interaction. Instead, we rely on a kinematic measurement of the bar length as an analogue to the dynamical length. The dynamical length should be less affected by the Scutum–Centaurus arm as we trace the signature of bar-supporting orbits instead of the density of stars. In our case, we use the mean radial fraction of the total velocity map, $\langle |v_R/v_{\mathrm{tot}}|\rangle$, which has a high value when the region is dominated by bar-supporting orbits as demonstrated in Section~\ref{sec::barlength}. Checking the N-body snapshots, which also have spiral arms connected to the bar, we also find a small signal (under the $\langle |v_R/v_{\mathrm{tot}}|\rangle=0.3$ threshold adopted above) in the $\langle |v_R/v_{\mathrm{tot}}|\rangle$ map in the spiral region, but it is insignificant compared to the signal from the bar stars. This indicates that the dynamics of stars in the spiral arm differ from that of disc stars but contribute mildly to the $\langle |v_R/v_{\mathrm{tot}}|\rangle$ value compared to the bar. The comparison we ran with the length estimated by the $x_1$ orbits, $R_{x_1}$, serves the same purpose, and shows that the $\langle |v_R/v_{\mathrm{tot}}|\rangle$ map can trace the length of the bar without much effect from the bar-spiral interaction. This kinematic analogue of the dynamical length is also a potential-independent measurement of the bar length. If we assume knowledge of the potential in the inner Galaxy as done in the dynamical analysis in Section~\ref{sec::orbital_family}, we can also classify the stars with $x_1$ orbits from our sample and use them to determine the bar length. Using the potential from \citet{Sormani_2022}, $R_{x_1}$ for the Milky Way is $3.8$~kpc, which is in agreement with its kinematic analogue $R_{\mathrm{b, kine}} = 4.0$~kpc. However, we do not further discuss the $R_{x_1}$ measurement of the Milky Way as it is model-dependent.

{We argued in Section~\ref{sec::barlength} that the $\langle |v_R/v_{\mathrm{tot}}|\rangle$ map traces the distribution of bar-supporting stars. Therefore, galactic bars comprised of different bar-supporting families could have different $\langle |v_R/v_{\mathrm{tot}}|\rangle$ maps. This is demonstrated by the top panels in Fig.~\ref{fig::bar_length_estimate}, in which the $\langle |v_R/v_{\mathrm{tot}}|\rangle$ maps have different shapes in Galaxy A and B because the orbital families are different. Galaxy A has orbital families as shown in Fig.~\ref{Appendix::SF_on_orbits} similar to the observation of the Milky Way bar; while Galaxy B has orbital families similar to the M2M model in \citet{Portail_2015}, in which over $95\%$ of stars have $\Omega_Z/\Omega_X\lesssim2$. However, despite the difference in the orbital family distribution, the dynamical length corresponds to $\langle |v_R/v_{\mathrm{tot}}|\rangle|\approx0.3$ for both galaxies. Therefore, testing on both Galaxy A and B reinforces the robustness of the method.

Estimating the dynamical length using the $\langle |v_R/v_{\mathrm{tot}}|\rangle$ map is similar to the method proposed by \citet{Petersen_2024}, which uses the $v_\perp$ map. However, as we argued in Section~\ref{sec::barlength}, our method is more suitable when the full 6D phase space of individual stars is measured, while the method in \citet{Petersen_2024} is better when the kinematic measurement is limited but has high signal-to-noise, which is better to implement on extragalactic observations. 


Now, we compare the kinematic analogue of dynamical length in this work, $R_{\mathrm{b,kine}} = 4.0$~kpc, to other previous dynamical length measurements of the Galactic bar. \citet{Lucey_2023} measures the "dynamical" length of the bar, defined by the apocentric radius of bar-supporting stars (not $x_1$ orbits), in various simulated galaxies. Combining the APOGEE and {\it Gaia} data, they integrate the observed Milky Way stars using the potential in the simulated galaxies with the same pattern speed of $\Omega_{\rm b}=41$~km s$^{-1}$ kpc$^{-1}$ and measure the same dynamical length for the Milky Way bar, and they find the simulated galaxy with the closest dynamical length as the best Milky Way surrogate. \citet{Lucey_2023} reports both a "dynamical" length $R_{\mathrm{freq}}=3.2$~kpc and $R_{x_1}=3.5$~kpc of that Milky Way surrogate galaxy as the bar length of the Milky Way (we use the same notation as \citet{Lucey_2023} for clarity). The $R_{x_1}$ measurement is still in broad agreement with our results, but the $R_{\mathrm{freq}}$ length is much shorter than our measurement. This could be because the pattern speed used in \citet{Lucey_2023} is faster than our measurement and also greater than the most recent measurement using resonances in the Galactic halo \citep{Dillamore_2024}, which would cause the bar length to be underestimated. Hence, the best-matched galaxies may also change if a different pattern speed is chosen. Also, in \citet{Lucey_2023}, a cut on $\Omega_Z/\Omega_X\lesssim3.3$ is applied when selecting bar-supporting stars. As we illustrate in Fig.~\ref{fig::orbital_family_data} \citep[also see][]{Portail_2015}, this effectively is a cut on the apocentre distribution of bar-supporting stars, and will cause the bar length to be underestimated. 

\citet{Vislosky_2024} compared the radial velocity field, $\overline{V_R}$ map, in {\it Gaia} DR3 with the cosmological and N-body simulations while taking the bar-spiral interaction into account. They find a consistent radial velocity field of galaxies that either have a short bar with a strong spiral arm or a long bar with a weak spiral feature. Our $R_{\mathrm{b,kine}}$ measurement is more consistent with their short bar model, which has $R_{\rm b}\sim 3.6~$kpc. Overall, the dynamical length measured in this work agrees with previous studies and is unique as a potential-independent measurement. 

\subsection{Pattern speed}

The pattern speed is estimated using the method developed in \citet{Dehnen_2023}, which builds on the continuity equation and is a 2D extension of the Tremaine-Weinberg (TW) method \citep{TW_1984}. \citet{Hilmi_2020} showed that the pattern speed estimation using the modified TW method in \citet{Sanders_2019} can lead to fluctuation up to $20\%$. However, \citet{Dehnen_2023} test the method with two sets of N-body simulations, and one of them (the "fiducial model" therein) has transient spiral arm features. The method successfully recovers the true pattern speed with a smaller fluctuation on the order of $1$~km s$^{-1}$ kpc$^{-1}$. This same is observed in out experiments. Many snapshots in our N-body simulation also show spiral features connected to the bar, and the pattern speeds recovered using the method do not deviate significantly from the true pattern speed calculated using the finite difference method. Hence, we argue the bar-spiral interaction does not have a strong effect on our measurement. Also, as the bar-spiral interaction scenario happens in our tested galaxies, any fluctuation caused by the method is included in the reported systematic uncertainty. 

The pattern speed measured and reportd here is $\Omega_{\rm b} = 34.1\pm2.4$~km s$^{-1}$ kpc$^{-1}$, in which the uncertainty includes both random and systematic error. This value is consistent with many recent measurements using a variety of different methods, including M2M and resonances features in the solar neighbourhood \citep{Clarke_2022, Binney_2020, Chiba_2021, Kawata_2021, Dillamore_2023, Dillamore_2024}. 

Resonance features in the solar neighbourhood provide important constraints on the bar pattern speed. \citet{Binney_2020} studied the kinematics in the solar neighbourhood and stars trapped by the bar resonance orbits and concluded a pattern speed of $\Omega_{\rm b}=35.2\pm1.0$~km s$^{-1}$ kpc$^{-1}$. \citet{Chiba_2021} used the metallicity gradient of stars trapped by the slowing-down bar and show that the pattern speed is $\Omega_{\rm b}=35.5\pm0.8$~km s$^{-1}$ kpc$^{-1}$. \citet{Kawata_2021} find the pattern speed of $34$ or $42$~km s$^{-1}$ kpc$^{-1}$ can both well-explain the local moving groups, in which the lower value is consistent with our measurements. Halo substructures can also be used as tracers of bar pattern speed. \citet{Dillamore_2023} first show that the bar resonances stars in the halo form constant energy ridges in the $E-L_z$ space. Combining with {\it Gaia} data, they find the ridges are consistent with the pattern speed of $35-40$~km\,s$^{-1}$ kpc$^{-1}$. Then, \citet{Dillamore_2024} extended the theory to $r-v_r$ space and explained the phase space chevron in \citet{Belokurov_2023} with bar resonances with a pattern speed of $35$~km s$^{-1}$ kpc$^{-1}$. Pattern speed measurements are also taken by directly observing the Galactic bar. \citet{Clarke_2019} compared data from the Variables in the Via Lactea (VVV) survey and {\it Gaia} DR2 with M2M models in \citet{Portail_2017}, qualitatively and found a pattern speed measurement of $37.5$~km s$^{-1}$ kpc$^{-1}$. Later, with quantitative analysis, the pattern speed measurement was revised down to $33.3\pm1.8$~km s$^{-1}$ kpc$^{-1}$ \citep{Clarke_2022}. All of these pattern speed estimates produced by studying both the Solar neighbourhood and the bar directly agree with the value measured using the continuity equation in this work. 

Note however that earlier implementations of the continuity equation resulted in pattern speeds higher than $\sim34$~km s$^{-1}$ kpc$^{-1}$ \citep{Sanders_2019, Bovy_2019, Leung_2023}. \citet{Sanders_2019} also showed that the pattern speed measurements become $31$~km s$^{-1}$ kpc$^{-1}$ when considering data on both sides of the bar, which suggested a $5-10$~km s$^{-1}$ kpc$^{-1}$ systematic error on the measurement. Also, the method used in \citet{Sanders_2019} is sensitive to the foreground and background perturbation because the continuity equation is integrated along the line-of-sight, which could be responsible for the fluctuation in the pattern speed measurements seen from the simulations in \citet{Hilmi_2020}. The spiral features are better handled in \citet{Dehnen_2023} because of the weighting function involved in the pattern speed calculation. \citet{Bovy_2019} used {\it Gaia} and APOGEE \texttt{AstroNN} distances to analyse the chemo-dynamic property of the Galactic bar. Applying the continuity equation with a pre-assumed exponential ellipsoidal density profile to the data produces a pattern speed of $41\pm3$~km s$^{-1}$ kpc$^{-1}$. The value was later revised to $40.1\pm1.8$~km s$^{-1}$ kpc$^{-1}$ in \citet{Leung_2023} with an updated dataset. One possible explanation for the discrepancy with our measurement could be the assumed density distribution. The suitability of the assumed ellipsoidal density distribution needs to be further tested. The difference could also be attributed to the incompleteness of observation due to the spectroscopic nature of the dataset. It is also shown in \citet{Bovy_2019} that the pattern speed measurements varied by $3-5$~km s$^{-1}$ kpc$^{-1}$ when changing the Sun's distance from the Galactic centre, the radial density gradient and the flattening of the ellipsoid. 

\subsection{Influence of a larger distance uncertainty}
\label{sec::discussion:larger_error}
As shown in Section~\ref{sec::data}, the median fractional distance uncertainty is $\sim10\%$, which is contributed by the intrinsic scatter of OSARG's PLR from \citet{OGLELMC_2007}, and the period uncertainty reported by the {\it Gaia} LPV catalogue \citep{GaiaDR3_LPV}. However, the intrinsic scatter of the PLR is still under debate. A different calibration method applied on the same dataset \citep[OGLEIII LMC,][]{OGLELMC_2007, OGLELMC_2009} yields different distance uncertainty (\citealt{OGLELMC_2007, Rau_2019, Hey_2023}; Zhang et al. in prep.). The parameterised PLR we adopted from \citet{OGLELMC_2007} showed intrinsic scatter that leads to less than $10\%$ distance uncertainty, which was calibrated using linear regression. \citet{Rau_2019} trained a machine learning model using the Random Forest \citep{Breiman_2001} and reported a similar scatter. \citet{Hey_2023} calibrated the LA-LPV distances using kernel density estimation, but the distance uncertainty reported is $\sim 15\%$. Zhang et al. (in prep.) use a similar method as that in \citet{Rau_2019} and show that the distance uncertainty is $\sim 10\%$ by comparing the calibrated luminosity distance to the globular cluster members with known distances. The distance uncertainties of the LA-LPV candidates in our sample are on the lower side of these measurements because we used the PLR from \citet{OGLELMC_2007}. The actual uncertainty could tend more towards $15\%$ as shown in \citet{Hey_2023}.

To verify the validity of the analysis and results, we perform similar tests to the bar length and the pattern speed as above but with $15\%$ distance uncertainty instead of $10\%$. The radial $\langle |v_R/v_{\mathrm{tot}}|\rangle$ profile along the bar major axis varies little when the distance uncertainty increases. The pattern speed estimation becomes more biased with larger distance uncertainty, and the estimated pattern speed is shown to be underestimated when the pattern speed is large, but a 1:1 consistency between the true and recovered pattern speed is still seen in most of the N-body simulation snapshots. Therefore, we argued that the results we obtained above are still valid even if the distance uncertainty is underestimated. The details are shown on Fig.~\ref{fig::appendix::bar_length} and \ref{fig::appendix::pattern_speed} in Appendix~\ref{Appendix::large_uncertainty}.

\section{Conclusions}
\label{sec::conclusion}
We use low-amplitude long period variables (LA-LPV, mainly OSARGs) as tracers of the kinematic and dynamic structures of the inner Milky Way. We use the LPV catalogue of \textit{Gaia} DR3 to select low-amplitude LPVs. We assign distance moduli of OSARGs using their period-luminosity relations calibrated in the Milky Way. We validate the assigned distances by comparing common stars in various samples, including {\it Gaia} geometric distances with good fractional parallax uncertainty, spectroscopic-based StarHorse distances, and globular cluster members. We demonstrate a good performance of the assigned distances with the median uncertainty of $\sim10\%$. We also compare the radial velocity uncertainties of the selected OSARGs to a randomly selected \textit{Gaia} DR3 sample and show that the radial velocity measurements of these LA-LPVs are reliable. We show the face-on view of our sample, which covers a large region in the Galactic disc and has considerable coverage on the far side of the Galaxy. Our sample covers heliocentric distances between $\sim1-18$~kpc and Galactocentric radii between $0-15$~kpc. Due to the photometric nature of the sample selection and the full-sky coverage of the \textit{Gaia} mission, there is only a weak spatial selection function. 

Our main results are:

\begin{enumerate}

\item We map the kinematic field of the inner Galaxy in terms of the mean radial, azimuthal and vertical velocity ($\overline{V_R}$, $\overline{V_\phi}$, $\overline{V_z}$), and radial velocity dispersion ($\sigma_R^\star$) in Fig.~\ref{fig::Kinematics_field_data}. The quadrupole, or butterfly pattern, in the $\overline{V_R}$ field is conspicuous confirming the existence of the bar in our sample. No systematic pattern is observed in the $\overline{V_z}$ map. 

\item We kinematically detect the Galactic bar using the mean radial fraction of the total velocity map, $\langle |v_R/v_{\mathrm{tot}}|\rangle$. We expect a high value of $\langle |v_R/v_{\mathrm{tot}}|\rangle$ in the bar region due to the elongated orbits of the bar-supporting stars. We observe an obvious signal in the $\langle |v_R/v_{\mathrm{tot}}|\rangle$ map in our sample corresponding to the shape of the bar and we verify the usefulness of this map with N-body simulations.

\item Imitating the observational uncertainty and selection function in a simulated galaxy, we find excellent consistency with our observed kinematic maps. This validates the quality of our sample. We find the heliocentric distance uncertainty biases the bar signature in the stellar density map and $\overline{V_R}$ map to align with the Sun-GC line, as expected \citep{Hey_2023, Vislosky_2024}. However, we also see that the orientation of the bar revealed by the $\langle |v_R/v_{\mathrm{tot}}|\rangle$ map is more robust against the heliocentric distance uncertainty. Hence, we use the signal in $\langle |v_R/v_{\mathrm{tot}}|\rangle$ to give a crude estimate of the bar angle of $25^\circ$, which is consistent with other studies \citep{Wegg_Gerhard_2013, Simion_2017, Clarke_2019, Bovy_2019}. 

\item We propose a potential-independent, purely kinematic method of measuring the dynamical length of the bar using the $\langle |v_R/v_{\mathrm{tot}}|\rangle$ map. We demonstrate that this map traces the orbital features of the bar-supporting stars and show that there exists a spatial match between the bar-supporting stars and the $\langle |v_R/v_{\mathrm{tot}}|\rangle$ map in those galaxies. More quantitatively, we find the bar length defined by $x_1$ orbits, $R_{x_1}$, corresponds to $\langle v_R/v_{\mathrm{tot}}\rangle |_{R_{b}}\approx0.3$. Using this feature, we estimate the kinematic analogue of the dynamical length of the bar as $R_{\mathrm{b,kine}}\sim4.0$~kpc.

\item We measure the bar pattern speed using the method developed in \citet{Dehnen_2023}, which was designed to extract the pattern speed from single simulation snapshots using the continuity equation. We demonstrate that the method also has good applicability to our data by running it on snapshots from N-body simulations after applying observational effects. Only small systematic biases and uncertainties are induced by the measurement errors and the selection function. The pattern speed obtained is $34.1\pm2.4$~km s$^{-1}$ kpc$^{-1}$, which is consistent with the values from many studies of the resonance features in the solar neighbourhood \citep{Binney_2020, Kawata_2021, Chiba_2021, Dillamore_2023, Dillamore_2024} and direct observation in the bar \citep{Clarke_2022}. 

\item We use the potential in \citet{Sormani_2022} to integrate the orbits and compute the orbital frequencies, $\Omega_i$. Selecting bar stars using their orbital frequencies ratio, we find $\sim 2,000$ bar-supporting stars in our sample. We study the vertical structure of orbits using the $\Omega_Z/\Omega_X$ ratio finding tht stars with greater $\Omega_Z/\Omega_X$ reside farther away from the centre in agreement with \citet{Portail_2015}.

\item We securely identify $\sim600$ $x_1$-orbital stars out of $2,000$ bar stars. Visually inspecting the rest of the selected stars, we find the majority also belong to the $x_1$ family tree. Moreover, $\sim90\%$ of bar stars have $\Omega_X/\Omega_Y\sim1$, hinting that they are $x_1$ orbit candidates. Hence, we conclude that the $x_1$ family constitutes the main building block of the Galactic bar.

\item Unlike the M2M model of the Galactic bulge built in \citet{Portail_2015}, we find an abundance of $x_1v_1$ banana orbits amongst our selected bar stars, suggesting they are the main constituents of the Galactic X-shape. This is in contrast to the predominance of the "brezel" orbits proposed by \citet{Portail_2015}. Plotting the spatial distribution of the banana and brezel orbits on the $X-Z$ plane, we see clear X-shaped structures; while banana orbits contribute more stars, brezel orbits give a sharper X-shape.

\end{enumerate}

For future work, we plan to include the metallicity measurements of these LA-LPV using \textit{Gaia} XP spectrum to expand the information to 7D or 8D. Then, a full chemo-dynamical analysis can be performed on this sample. A preliminary result suggests our LA-LPV sample covers a large range of metallicity from $-1.5$ to $0.5$ dex.

\section*{Data availability}

\textit{Gaia} data used in this work is publicly available. The raw \textit{Gaia} LPV catalogue used in this work can be downloaded from the Gaia archive at \url{https://gea.esac.esa.int/archive/}, and the ADQL query code used is attached in Appendix~\ref{ADQL}. The sample constructed in this work is availible at \url{https://zenodo.org/records/13285574}. 

\section*{Acknowledgements}

We thank the reviewer for the helpful comments. We thank Adam Dillamore and Elliot Davies for their inspirational discussions and for their help in setting up the N-body simulations. We thank Eugene Vasiliev for helpful comments on the draft of the paper

HZ thanks the Science and Technology Facilities Council (STFC) for a PhD studentship. 
VB acknowledges support from the Leverhulme Research Project Grant RPG-2021-205: "The Faint Universe Made Visible with Machine Learning". SK thanks the Marshall Scholarship for her PhD funding. JLS acknowledges support from the Royal Society (URF\textbackslash R1\textbackslash191555).

This work has made use of data from the European Space Agency (ESA) mission
{\textit{Gaia}} (\url{https://www.cosmos.esa.int/gaia}), processed by the {\textit{Gaia}}
Data Processing and Analysis Consortium (DPAC,
\url{https://www.cosmos.esa.int/web/gaia/dpac/consortium}). Funding for the DPAC
has been provided by national institutions, in particular the institutions
participating in the {\textit{Gaia}} Multilateral Agreement.





\bibliographystyle{mnras}
\bibliography{bibliography} 

\appendix

\section{Effect of selection function on the orbital family distribution}
\label{Appendix::SF_on_orbits}
Although we show that the selection function (SF) affects our kinematic results only weakly, the significance of the SF effects on the dynamics is also worth testing. To achieve this, we employ the N-body simulated Galaxy A because it has a similar $\Omega_Z/\Omega_X$ distribution to the observations. We use the same techniques of applying the selection function to the simulation by matching the simulated particles to observed stars in the spatial distribution. Matching every observed star to a simulated star that is closest to it in $x-y-z$ space, we drop all simulated particles that failed to become the closest match to observation or for which the closest matched distance is greater than $0.5$~kpc. After this procedure, the number of simulated particles is roughly the same as the observed number of stars, and the spatial distribution is almost identical. Selecting bar-supporting stars using the same method in Section~\ref{sec::dynamics}, the orbital distribution is shown in Fig.~\ref{fig::appendix::SF_orbits}, in which the red line shows the raw distribution before applying SF, and the black is the distribution after applying SF to Galaxy A.

As we mentioned in Section~\ref{sec::data}, the main selection effects in our sample are that 1) the inner $\sim150$~pc of the Galaxy is missing due to dust extinction and 2) the far side of the disc is not mapped as completely as the near side. This introduces two effects that could potentially affect our conclusion. The banana orbits at $\Omega_Z/\Omega_X=2$ are slightly enhanced since they stay out of the galactic plane more than the other families. The orbits with $\Omega_Z/\Omega_X\gtrsim2.5$ are slightly depleted because they tend to extend farther away from the galactic centre, and therefore, they are mainly observed on the near-side of the disc, while the families with smaller $\Omega_Z/\Omega_X$ are observed on both sides. However, as evident in Fig.~\ref{fig::appendix::SF_orbits}, both effects are small.  We check the SF effects on the $\Omega_X/\Omega_Y$ distribution and find the changes are also insignificant. Combining with the minor effects from the uncertainty as we see in Fig.~\ref{fig::orbital_family_data}, we therefore argue that the orbital family distribution we find in Section~\ref{sec::dynamics} is close to the real distribution in the Milky Way bar. 

\begin{figure}
    \centering
    \includegraphics[width = \columnwidth]{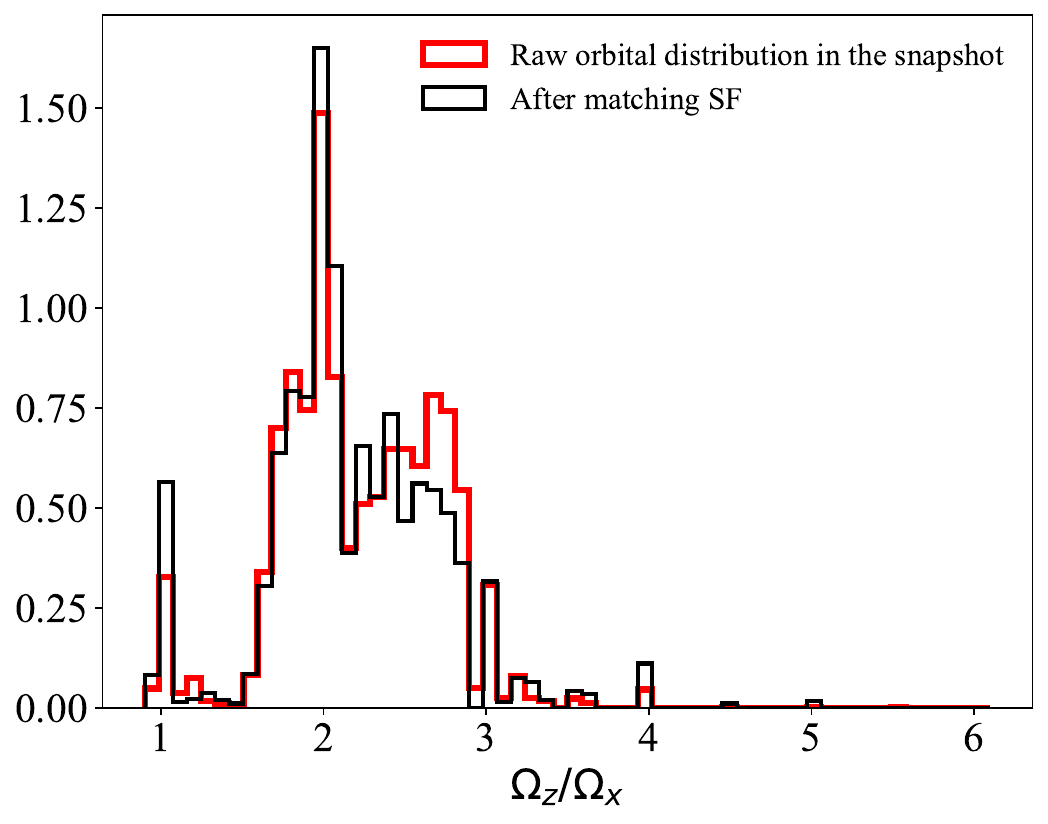}
    \caption{The effect of the selection function on orbital distribution is shown in this plot. The red and black lines are the orbital families before and after applying the selection function to Galaxy A, respectively.}
    \label{fig::appendix::SF_orbits}
\end{figure}

\section{Side-on projection of bar orbits}
\label{Appendix::side_on_projection}
The Galactic bar is composed of many orbital families classified by the $\Omega_Z/\Omega_X$ value. Together, they assemble a boxy/peanut (BP) morphology with an X-shape structure embedded in an overall ellipsoidal density distribution. In Fig.~\ref{fig::appendix::side_on_proj}, we present the side-on projection ($X-Z$ plane) of the selected bar stars in our sample in each $\Omega_Z/\Omega_X$ bin. The background is the density distribution by stacking the orbits of stars in each $\Omega_Z/\Omega_X$ bin, which effectively represents the bar density contributed by this orbital family. Stars with smaller $\Omega_Z/\Omega_X$ have more X-like shapes, while stars with larger $\Omega_Z/\Omega_X$ are more elongated and boxy-like. 

\begin{figure}
    \includegraphics[width = \columnwidth]{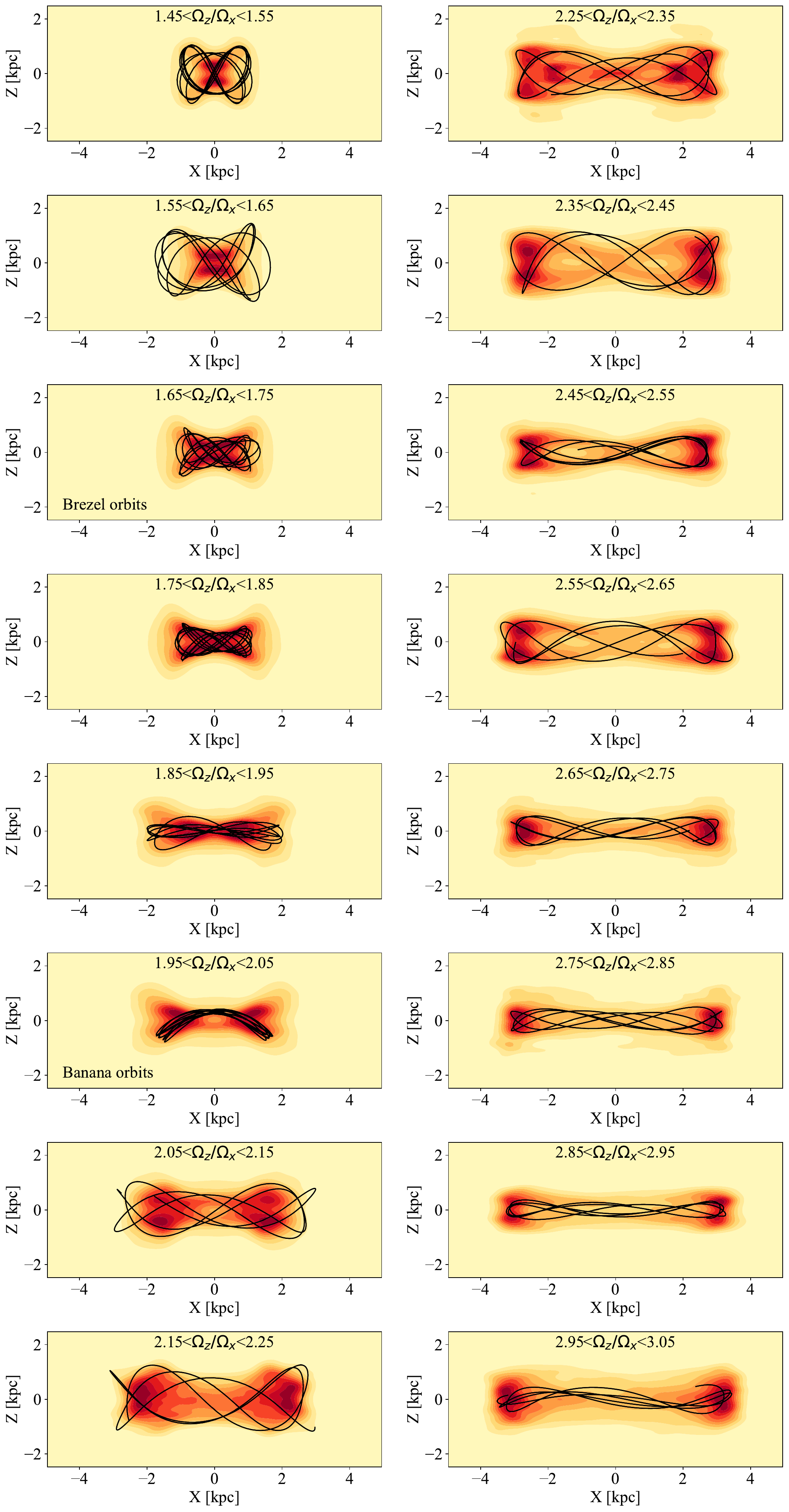}
    \caption{The side-on projection of orbital families in the Galactic bar. Stars are binned into $\Omega_Z/\Omega_X$ segments as denoted in each panel. The background shows the density contributed by this orbital family by stacking the integrated orbits together. The overlaid line is a selected representative example of orbits in this family.}
    \label{fig::appendix::side_on_proj}
\end{figure}

\section{Bar length and pattern speed measurement with $15\%$ distance uncertainty}

\label{Appendix::large_uncertainty}
\begin{figure*}
    \centering
    \includegraphics[width = 0.9\textwidth]{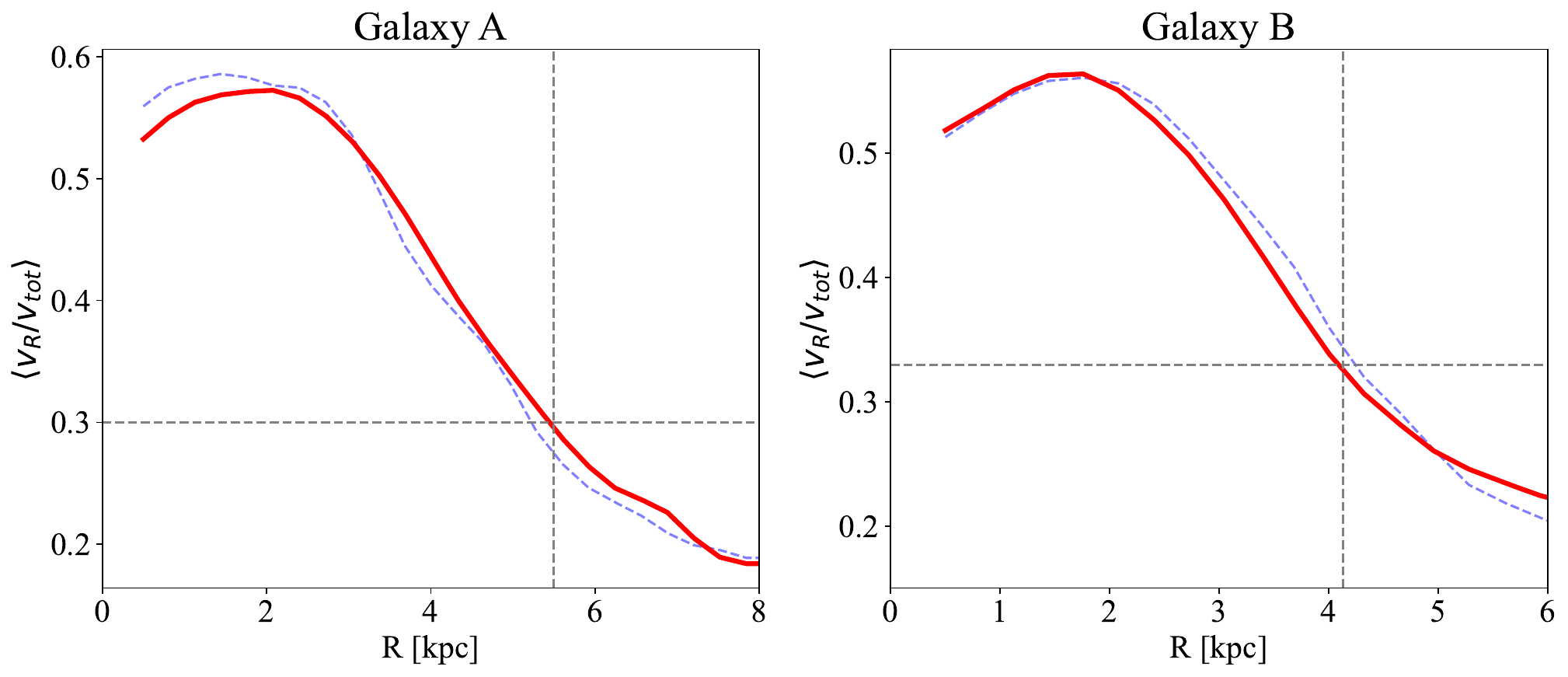}
    \caption{The radial $\langle |v_R/v_{\mathrm{tot}}|\rangle$ along the bar major axis as a function of radius in Galaxy A (left) and B (right). The red line is the radial profile of the raw snapshot, and the blue dashed line is the radial profile after applying $15\%$ distance uncertainty and the selection function.}
    \label{fig::appendix::bar_length}
\end{figure*}

\begin{figure}
    \centering
    \includegraphics[width = 0.95\columnwidth]{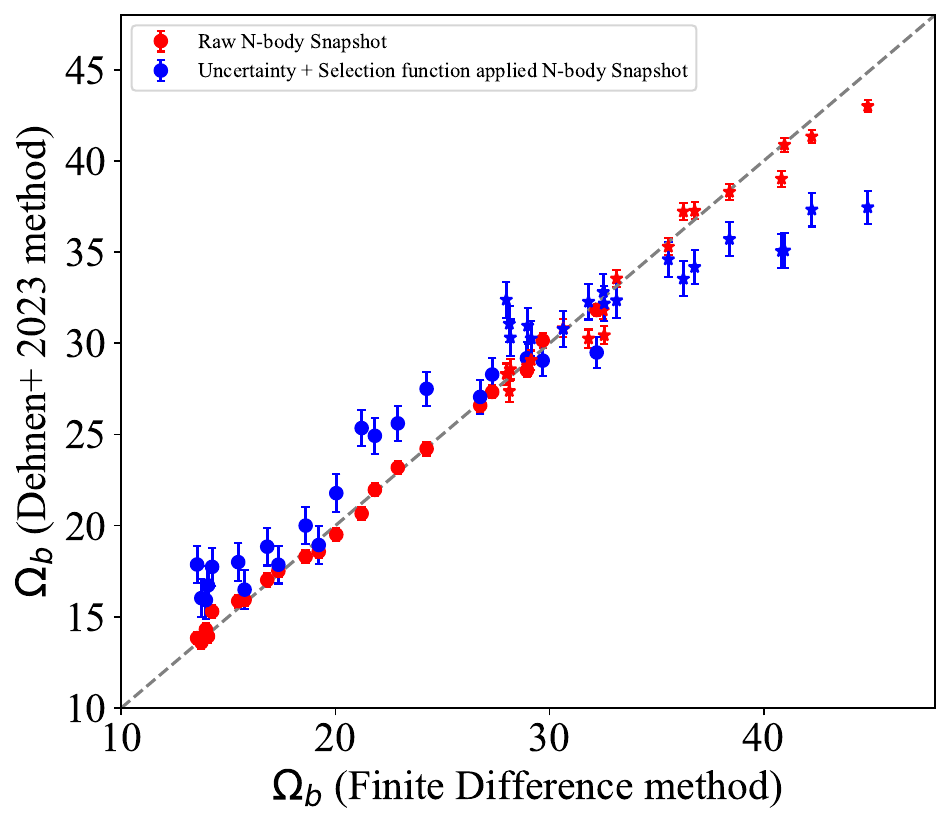}
    \caption{The recovered pattern speed compare to the true pattern speed for different N-body time snapshots. The red dots is the pattern speed estimate by applying the method in \citet{Dehnen_2023} to the raw N-body snapshots, and the blue is that after applying $15\%$ distance uncertainty and the selection function. The grey dashed line is the 1:1 line.}
    \label{fig::appendix::pattern_speed}
\end{figure}

As we discussed in Section~\ref{sec::discussion:larger_error}, the heliocentric distance uncertainty could be underestimated in our sample, in which the fractional distance error could be as large as $15\%$. Therefore, we repeat the same analysis in Section~\ref{sec::barlength} and \ref{sec::patternspeed_measurement} with $15\%$ distance uncertainty. The blue dashed line in Fig.~\ref{fig::appendix::bar_length} is the radial $\langle |v_R/v_{\mathrm{tot}}|\rangle$ value along the major axis of the bar after applying $15\%$ distance uncertainty and the selection function. The difference between the radial profile before and after applying the observational caveats are still small. 

Similar to Fig.~\ref{fig::test_dehnen2023_method}, we show the bias on the pattern speed recovery caused by the observational uncertainty and selection function in Fig.~\ref{fig::appendix::pattern_speed} but with a $15\%$ distance error. The bias is larger with the larger distance uncertainty, and the pattern speed is also consistently underestimated at high pattern speeds. However, the true and recovered pattern speeds still follow the 1:1 line, so we argue that the pattern speed estimation using the method in \citet{Dehnen_2023} is still valid with the presence of $15\%$ distance uncertainties and the incompleteness. 

\section{ADQL Query}
\label{ADQL}
We attach below the ADQL query code for reproducing the raw catalogue that we used to select the LA-LPV candidates from the \textit{Gaia} archive. 


\begin{verbatim}
SELECT
gs.source_id,gs.ra,gs.dec,
lpv.*, vari.best_class_score, 
vari.best_class_name, tmass.*
FROM gaiadr3.gaia_source AS gs
JOIN gaiadr3.vari_long_period_variable AS lpv 
ON gs.source_id = lpv.source_id
JOIN gaiadr3.vari_classifier_result as vari 
ON gs.source_id = vari.source_id
JOIN gaiadr3.tmass_psc_xsc_best_neighbour AS xm
ON gs.source_id = xm.source_id
JOIN gaiadr3.tmass_psc_xsc_join AS xj
ON xm.clean_tmass_psc_xsc_oid = 
xj.clean_tmass_psc_xsc_oid
JOIN gaiadr1.tmass_original_valid AS tmass 
ON xj.original_psc_source_id = tmass.designation
WHERE vari.best_class_score > 0.8
\end{verbatim}








\bsp	
\label{lastpage}
\end{document}